\newif\ifanonymized
\anonymizedfalse     

\newif\ifsubmit
\submittrue	

 \documentclass[acmlarge,screen]{acmart}
 \hypersetup{
  colorlinks=true,
  linkcolor=blue,     
  citecolor=blue,     
  urlcolor=blue       
}

\usepackage{fancyhdr}

\usepackage{url}

\usepackage{color}

\usepackage[mmddyyyy,hhmmss]{datetime}

\usepackage{xspace}

\usepackage{graphicx}
\DeclareGraphicsExtensions{.pdf,.png,.eps,.ps,.jpg}

\ifsubmit
  \newcommand{\hey}[1]{\relax}
  \newcommand{\heyvarun}[1]{\relax}
  \newcommand{\bibtex}[1]{\relax}
  \newcommand{\note}[1]{\relax}
\else
  \newcommand{\hey}[1]{\textcolor{magenta}{[{#1}]}}
  \newcommand{\heyvarun}[1]{\textcolor{blue}{[Varun: {#1}]}}
  \newcommand{\note}[1]{\par\textcolor{magenta}{Note: {#1}}\par}
  \newcommand{\bibtex}[1]{\textcolor{red}{@bibtex}\{#1\}} 
\fi

\newcommand{\hide}[1]{\relax}

\newcommand{\seclabel}[1]{\label{sec:#1}}

\usepackage{longtable}
\usepackage{tabularx}
\newcommand{\jiachen}[1]{\textcolor{black}{#1}}
\usepackage{color, colortbl}
\usepackage{anyfontsize}
\usepackage{xcolor}
\definecolor{DarkGreen}{HTML}{5DAC81}
\usepackage{etoolbox}
\usepackage{textcase} 

\makeatletter
\patchcmd{\thebibliography}
  {\list}
  {\interlinepenalty=10000\list}
  {}{}
\makeatother

\pdfoutput=1

\begin{document}

\setcopyright{cc}
\setcctype{by-nc-sa}
\acmJournal{IMWUT}
\acmYear{2025} \acmVolume{9} \acmNumber{3} \acmArticle{101} \acmMonth{9} \acmPrice{}\acmDOI{10.1145/3749508}



\title[Vital Insight: Assisting Experts’ Context-Driven Sensemaking of Multi-modal Personal Tracking Data]{Vital Insight: Assisting Experts' Context-Driven Sensemaking of Multi-modal Personal Tracking Data Using Visualization and Human-in-the-Loop LLM}




\author{Jiachen Li}
\authornote{Corresponding authors}
\orcid{0000-0002-6084-5131}
\email{li.jiachen4@northeastern.edu}
\affiliation{%
  \institution{Northeastern University}
  \city{Boston}
  \state{Massachusetts}
  \country{USA}
}

\author{Xiwen Li}
\orcid{0009-0008-3602-0392}
\email{li.xiwe@northeastern.edu}
\affiliation{%
  \institution{Northeastern University}
  \city{Boston}
  \state{Massachusetts}
  \country{USA}
}

\author{Justin Steinberg}
\orcid{0009-0003-9880-0021}
\email{steinberg.ju@northeastern.edu}
\affiliation{%
  \institution{Northeastern University}
  \city{Boston}
  \state{Massachusetts}
  \country{USA}
}

\author{Akshat Choube}
\orcid{0000-0002-7969-322X}
\email{choube.a@northeastern.edu}
\affiliation{%
  \institution{Northeastern University}
  \city{Boston}
  \state{Massachusetts}
  \country{USA}
}

\author{Bingsheng Yao}
\orcid{0009-0004-8329-4610}
\email{b.yao@northeastern.edu}
\affiliation{%
  \institution{Northeastern University}
  \city{Boston}
  \state{Massachusetts}
  \country{USA}
}

\author{Xuhai Xu}
\orcid{0000-0001-5930-3899}
\email{xx2489@columbia.edu}
\affiliation{%
  \institution{Columbia University}
  \city{New York}
  \state{New York}
  \country{USA}
}

\author{Dakuo Wang}
\orcid{0000-0001-9371-9441}
\email{d.wang@northeastern.edu}
\affiliation{%
  \institution{Northeastern University}
  \city{Boston}
  \state{Massachusetts}
  \country{USA}
}

\author{Elizabeth Mynatt}
\orcid{0000-0001-8486-9384}
\email{e.mynatt@northeastern.edu}
\affiliation{%
  \institution{Northeastern University}
  \city{Boston}
  \state{Massachusetts}
  \country{USA}
}

\author{Varun Mishra}
\orcid{0000-0003-3891-5460}
\email{v.mishra@northeastern.edu}
\affiliation{%
  \institution{Northeastern University}
  \city{Boston}
  \state{Massachusetts}
  \country{USA}
}

\renewcommand{\shortauthors}{}





\begin{abstract}
Passive tracking methods, such as phone and wearable sensing, have become dominant in monitoring human behaviors in modern ubiquitous computing studies. 
While there have been significant advances in machine-learning approaches to translate periods of raw sensor data to model momentary behaviors, (e.g., physical activity recognition), there still remains a significant gap in the translation of these sensing streams into meaningful, high-level, context-aware insights that are required for various applications (e.g., summarizing an individual's daily routine). 
To bridge this gap, experts often need to employ a context-driven sensemaking process in real-world studies to derive insights.
For instance, current approaches in the field can reliably predict ``walking outdoors'' by contextualizing  accelerometer and GPS data. Sensemaking, however, involves being able to notice patterns of periodic ``walking'' and ``stationary'' events, and even infer ``walking the dog'' after realizing the alignment with regular routines reported through surveys or self-reports. 
This process often requires manual effort and can be challenging even for experienced researchers due to the complexity of human behaviors. 

We conducted three rounds of user studies with 21 experts to explore solutions to address challenges with sensemaking.
We follow a human-centered design process to identify needs and design, iterate, build, and evaluate Vital Insight (VI), a novel, LLM-assisted, prototype system to enable human-in-the-loop inference (sensemaking) and visualizations of multi-modal passive sensing data from smartphones and wearables.
Using the prototype as a technology probe, we observe experts' interactions with it and develop an expert sensemaking model that explains how experts move between direct data representations and AI-supported inferences to explore, question, and validate insights. 
Through this iterative process, we also synthesize and discuss a list of design implications for the design of future AI-augmented visualization systems to better assist experts' sensemaking processes in multi-modal health sensing data.

\end{abstract}


\maketitle  





\section{INTRODUCTION} 
\seclabel{introduction}
%
The ubiquitous presence of sensor-rich smartphones and wearables has prompted researchers to use data from these devices to track living activities for various outcomes including health tracking, context-aware applications, and digital health interventions~\cite{friedewald2011ubiquitous,orwat2008towards,mishra2018investigating,ho2005using,kunzler2019exploring,boateng2019experience,farber2011aging,li2023privacy,kononova2019use,vargemidis2020wearable,seifert2017use,10.1145/3706598.3714086}. 
However, longstanding socio-technical gaps persist in collecting data and generating valuable insights, due to challenges such as limited information, noisy data, ambiguous contexts, and variations in individual behavior~\cite{ackerman2000intellectual,chen2002context,vcas2011ubiquitous,edwards2001home,bourguet2011uncertainty,kumar2009challenges}.
Most efforts in this area focus on developing algorithms to model windows of raw sensor data into labels of activity, behavior, or outcomes, such as predicting physical activity, stress, and depression~\cite{10466375,amin2023investigating,mishra2020continuous,cornacchia2016survey,doherty2014tracking,wang2018tracking}.
There is however, a lack of exploration in generating high-level insights about these periodic behaviors such as ``is it a normal day'' or a summary of someone’s day that is often more valuable for various stakeholders~\cite{mynatt2001digital,adler2024beyond}.
In real-world deployment, researchers often have to find alternate ways to derive insights that are needed for practical applications.

Currently, the gold standard to derive high-level insights is still a manual examination of sensing data by experts, often referred to as \textbf{context-driven sensemaking}—an iterative procedure where experts alternate between data and develop an appropriate understanding of human behavior situated in specific contexts\cite{faisal2009classification,adler2024beyond,bietz2016opportunities,boman2015sensemaking,jones2016sensemaking}.  
In real-world studies, experts synthesize context-driven inferences with well-reasoned explanations, fill in the gap in data, and identify needed actions.
For instance, experts may first notice a relatively low and stable heart rate, along with a sedentary period around 3 p.m., and then attempt to find a possible explanation by examining other modalities. 
They might notice a long period of no phone usage, deduce that the person is likely resting based on their older age, and later confirm this inference by reviewing a conversation with the chatbot in the morning in which the person mentioned getting up early and planning to a nap in the afternoon.
These sensemaking processes allow experts to generate high-level insights and valuable context to explain the raw sensing data and even validate machine learning results.

Although experts' sensemaking efforts are crucial, the process remains fraught with uncertainty and challenges, even for those with expertise~\cite{boman2015sensemaking,jones2016sensemaking}.
Past studies have attempted to scaffold the human sensemaking process, identifying various sensemaking modes and schemes~\cite{russell1993cost, pirolli2005sensemaking, klein2006making, krizan1999intelligence, barrett2009interpretation, zhang2014towards}. 
While these classic paradigms are valuable, they are often too general and insufficient to inform the design of assistive systems in this specific scenario of making sense of personal tracking data by experts. 
This specific sensemaking process is a complex and mentally demanding task that requires experts to understand not only the technical aspects of the sensing technologies but also the domain-specific knowledge about human behavior and real-world commonsense. 
The unique challenges and needs of experts' sensemaking process with multi-modal personal tracking data remain underexplored in current studies, which is the focus of our research.

Other than identifying the challenges, we are also curious about how we might design technology to assist experts' sensemaking processes. 
Since this scenario is often neglected by past research, there is also a lack of tool that targets to assist the specific sensemkaing process.
Some prior works have used techniques such as visualizations and activity recognition algorithms to provide basic assistance, however, often focus narrowly on presenting direct data representations instead of supporting the more complex sensemaking tasks experts undertake in real-world practice~\cite{ponnada2021signaligner, hagen2023annots,barz2016multimodal, martindale2018smart, diete2017smart}. 
With the emergence of AI, such as Large Language Models (LLMs), researchers have begun to recognize LLMs' capability to memorize and use 'common sense' and world knowledge to directly generate insights from sensor data, such as identifying activities, making diagnostic predictions, and more~\cite{englhardt2024classification, yang2024drhouse, ferrara2024large, ji2024hargpt, fang2024physiollm,king2024sasha}.
In many health-related scenarios, there has been a long-standing trust issue with experts interacting with AI-generated insights~\cite{choudhury2024large, kerasidou2022before, ryan2020ai}. 
However, there is also a growing body of work that has achieved success by designing with respect to human workflows~\cite{yang2024talk2care, 10.1145/3613904.3642343}.
Realizing this potential, we want to further explore whether AI can be leveraged as a form of assistance in the scenario of complex sensemaking of multi-modal personal tracking data by experts, and if so, how they should be designed to better fit into the sensemaking procedures.


In this paper, we present a user-centered design process to explore the complex sensemaking process of experts when interpreting multi-modal health sensing data and the opportunity to support this process. 
First, we conduct formative interviews with 12 experts\footnote{We define ``experts'' as people who are experienced researchers in ubiquitous computing who have utilized passive sensing data for health-related applications. We will discuss in detail in Section 3.1} to answer \textbf{RQ1: What are the challenges and needs when `making sense' of personal tracking data in current practices?} (Study 1). 
Guided by our findings, we develop an initial prototype, Vital Insight version 1 (VI-1), that leveraged both visualization and Large Language Models to assist experts' sensemaking. 
We then conduct an exploratory user testing session with 13 experts using VI-1 as a technology probe to perform several sensemaking tasks (Study 2). 
Based on the user testing and feedback on VI-1, we iterate on our prototype and designed Vital Insight version 2 (VI-2), mainly refining the LLM augmentation to include more human-in-the-loop guidance.
To further evaluate the modified AI components, we conduct an A/B testing session with 20 experts to investigate whether the AI augmentation is indeed helpful in their sensemaking process (Study 3). 
Our findings from the two studies indicate that the experts found VI to be a valuable prototype for assisting them in conducting evidence-based sensemaking, giving high ratings in dimensions such as trustworthiness and helpfulness.
Experts also found their sensemaking tasks more satisfying with AI components than without them.

Through these three user studies, we analyze the experts' interactions with the prototype to answer \textbf{RQ2: What is the sensemaking process experts employ when interpreting personal tracking data?}. 
We identify the sensemaking process in which experts move from sensing data and contextual information to six types of patterns, personas, and actions. 
We also discuss the potential of AI at each step, along with a high-level sensemaking model that outlines the interaction between experts, direct data representation, and AI-supported inference.
Furthermore, we synthesize and refine a final list of design implications to answer \textbf{RQ3: What are the design implications for AI-augmented visualization systems aimed at assisting the sensemaking of personal tracking data?} 
Given our iterative approach, the findings from the formative interviews significantly inform our user study and evaluations; thus, we report the methods separately in their respective sections.

\medskip
\noindent
In this paper, we make four important \textbf{contributions}
\begin{itemize}
    \item Identify the \textbf{challenging sensemaking scenarios} in experts' practices to understand personal tracking data
    \item \jiachen{Construct} the \textbf{sensemaking model} of experts when dealing with complex personal tracking data
    \item Synthesize a list of \textbf{design implications} of the AI-assisted visualization system in helping experts conduct sensemaking of personal tracking data
\end{itemize}

\section{BACKGROUND} 
\seclabel{background}

\subsection{From Raw Data to Valuable Insights: Experts' Sensemaking Model}
The concept of sensemaking emerged in the late 1970s~\cite{dervin1983overview, norman1975role} and became more prominent in the 1990s, particularly in research related to organizations, education, and decision-making~\cite{dervin1992mind, greenberg1995blue, klein2006making}. Since then, several research fields have adopted and expanded upon this notion, developing models tailored to explain the sensemaking process within their specific domains~\cite{russell1993cost, pirolli2005sensemaking,klein2006making, krizan1999intelligence, barrett2009interpretation, zhang2014towards}. Despite variations in definitions, the overarching idea remains that sensemaking is an iterative process involving information search and developing an understanding of a situation to take appropriate action. Russel et al. did multiple case studies to design the cost structure of the sensemaking process and presented a schema-based iterative model~\cite{russell1993cost}. Pirolli and Card studied the sensemaking process in the context of intelligence analysts and proposed a model having information foraging and sensemaking loops ~\cite{pirolli2005sensemaking}. Similarly, Klein et al.~\cite{klein2007data} presented a data frame model where sensemaking involves a cyclic process of aligning data with frames (explanatory structures) and adapting the frame to data.

Some past works have also investigated sensemaking within the context of passive sensing for personal health. Mamykina et al.~\cite{mamykina2017personal}, situated self-tracking sensemaking behaviors of adolescents with Diabetes and their caretakers with the existing literature and proposed a theoretical framework for these behaviors. Co\c{s}kun and Karahano\u{g}lu~\cite{cocskun2023data} did a systematic literature review of sensemaking practices in HCI literature and defined four sensemaking modes self-trackers go through (i.e., self-calibration, data augmentation, data handling, and realization). Researchers have also studied personal health sensemaking in collaborative online platforms~\cite{puussaar2017enhancing, figueiredo2020patient} and information delivery~\cite{calvillo2011pamphlet}. Although researchers have investigated sensemaking in self-tracking, sensemaking of participants’ data by experts remains underexplored. Sensemaking by experts differs from that of participants, as it generally involves more comprehensive analysis and often lacks the contextual information that participants possess. In this paper, we fill this gap by understanding experts' sensemaking process through user studies and the Vital Insight prototype.

\subsection{Tools to Assist the Sensemaking of Personal Tracking Data}
Some past research has acknowledged the challenges of making sense of personal tracking data and has developed various tools to assist with different tasks. 
Among those tools aimed at experts, many focus on specific tasks, such as activity annotation on accelerometer data~\cite{ponnada2021signaligner, hagen2023annots}, multi-modal sensor data~\cite{barz2016multimodal, martindale2018smart, diete2017smart}, video analysis~\cite{10.1145/2968219.2971461, diete2017recognizing,sztyler2016body}, and spatial-temporal events~\cite{rawassizadeh2019indexing}. 
The majority of these tools are built around visualization, where the system serves as an interactive interface to better display data and assist in the annotation process. 
Beyond activity annotation, some tools take a broader approach to understanding multi-modal data. 
For instance, Lifestream integrates different data modalities and features into one dashboard to facilitate the exploration and evaluation of personal data streams with users~\cite{hsieh2013lifestreams}. 
Chung et al. used the EventFlow tool, which aggregates workflow information to highlight patterns and variabilities, helping different stakeholders understand the daily routines of older adults~\cite{chung2017examining}. 
Some works also explored other modalities, such as conversational interfaces in helping the understanding of data~\cite{kim2021data,shen2022towards,hearst2019would,aurisano2015show,10.1145/3706598.3714272}.

\jiachen{In addition to visualizing routine/activity-related personal tracking data, there is a substantial body of work in the visualization community focused on time-series and multi-modal data. Kehrer et al. highlight that multi-modal scientific data often involves different acquisition modalities, varying spatial/temporal resolutions, and diverse data types~\cite{kehrer2012visualization}. These differences create challenges in visualization design, particularly in aligning data for effective fusion and comparative analysis. One common solution is small multiples, a technique where various attributes are plotted along a consistent x-axis scale, which supports multifaceted exploration~\cite{bavoil2005vistrails}, pattern recognition~\cite{lekschas2017hipiler}, and data exploration~\cite{van2013small}. This approach is especially useful for tasks that require direct visual comparisons of time-series data, aiding in the comparison, exploration, and analysis of trends~\cite{javed2010graphical,boyandin2012qualitative,bach2015small,meyer2010pathline,heer2009sizing}. Researchers have also investigated specific visualization methods for time-series data, such as line graphs for trends, bar or circle charts for cumulative or cyclic data~\cite{matthews1997worm,harris1999information}, and axis-based techniques, along with parallel coordinates for multivariate data~\cite{10.1145/967900.968153,970ebe970164479f81ed8be97a37f73e,9362264}.
In addition to crafting the visual representation of data, past work has also proposed principles for the interaction and customization with visualizations. For instance, Schneiderman’s mantra ``Overview first, zoom and filter, then details-on- demand'' describes how data should be presented on
screen so that it is most effective for users~\cite{shneiderman2003eyes}.
Other visual interaction methods proposed by past works include explore, reconfigure, select, encode, abstract, elaborate, filter, and connect.~\cite{yi2007toward,blascheck2018exploration,fang2020survey}. 
}

Most of these tools mentioned above focus on better data representation as a means to help experts and users understand and reflect on the data. 
However, some tools recognize that simply presenting data is not enough for meaningful understanding, and leverage AI to provide automatic assistance. 
For example, machine learning models have been widely used to attach labels to sensing data and have proven successful in tasks such as activity recognition, stress detection and mental health~\cite{mishra2020evaluating,mishra2020continuous,10.1145/3351274}, sleep analysis~\cite{kristiansen2021machine} and more.
However, most of these tools are limited to ground truth labels and can only perform basic recognition tasks, failing to assist in more complex levels of understanding of personal tracking data. 
While these labels serve as good recognition results, they fall short in supporting experts in their sensemaking process.
With the emergence of large language models (LLMs), researchers have begun adopting their ``common sense'' knowledge to provide higher-level inferences on sensor data~\cite{cosentino2024towards,fang2024physiollm,ji2024hargpt,ouyang2024llmsense,kim2024health}. 
\jiachen{For instance, HARGPT uses LLMs to identify activities from raw IMU data~\cite{ji2024hargpt}; Health-LLM evaluated zero/few-shot learning and fine-tuning on sensor data~\cite{kim2024health}; and LLMSense and PhysioLLM use text-formatted sensor data and prompts to derive high-level inferences from sensor traces~\cite{ouyang2024llmsense,fang2024physiollm}.}
However, these tools are often designed not to assist experts but to compare their accuracy with experts' results. 
Due to issues like hallucinations, experts still struggle to trust and validate these results, making manual collaboration with such systems challenging~\cite{lu2023human,chiang2023can}.
\jiachen{Some recent works have begun to explore the use of AI as an aid to human sensemaking in fields such as genomics~\cite{mastrianni2024aienhancedsensemakingexploringdesign}, social relationships~\cite{10.1145/3411764.3445290}, annotation~\cite{10.1145/3411763.3451798}, stroke diagnosis~\cite{jussupow2022radiologists}, supply chain survival~\cite{sharma2025artificial}, and more. Wenskovitch et al. further categorize AI’s roles in data sensemaking into four types: explorer, investigator, teacher, and judge, emphasizing the need for mutually intelligible, bidirectional communication~\cite{9619901}.}
\jiachen{With the increasing focus on using AI as an assistance in the human sensemaking process,} we recognize that there is still a lack of tools that leverage both evidence and helpful inferences to support experts in interpreting personal tracking data.

\section{STUDY 1: FORMATIVE INTERVIEW}
\label{sec:interview}
In this section, we describe our initial interviews with experts about the challenges they face in deriving insights from personal tracking data, highlighting the needs and opportunities in assisting experts in their sensemaking process. 

\subsection{Method}
We conducted IRB-approved formative interviews with 12 experts in multi-modal sensing data with a \$15 compensation. 
The inclusion criteria for experts are experienced researchers in ubiquitous computing who have utilized passive sensing data for health-related applications. 
In addition to self-selection, we also added objective inclusion criteria, requiring publication in IMWUT, CHI, JMIR, and other relevant conferences or journals.
Through these hour-long semi-structured online interviews, we wanted to better understand the experts' unique research processes while inferring complex real-world personal tracking data.
After obtaining verbal consent, we recorded the audio and transcripts of the interviews. 
Two members of the research team carefully reviewed the transcripts and notes, and independently coded specific segments of the data, employing an inductive open coding approach~\cite{corbin1990grounded}.
The coders then exchanged codebooks and discussed to harmonize and merge the codes by consensus, and conducted a thematic analysis under the principles of Grounded Theory~\cite{braun2006using, glaser1968discovery}. 
We recruited a total of 12 experts who had various experiences in working with personal tracking data (Table \ref{tab:demo}), reached data saturation and stopped further recruitment~\cite{Guest2006HowMI,Ando2014AchievingSI}. 

\begin{table}[t]
\caption{Demographics of experts participating in the three user studies.}
\small
\begin{tabular}{@{}ccccc@{}}
\hline\hline
Participant & Gender & Education & \begin{tabular}[c]{@{}c@{}}Years of Experience\end{tabular} & \begin{tabular}[c]{@{}c@{}}Participation in Study\end{tabular} \\
\hline
P1  & M  & Doctorate & 8 years & 2,3 \\
P2 & F  & Doctorate & 1 year & 1,2  \\
P3  & F & Master & 3 years & 1,2,3\\
P4  & F  & Bachelor & 4 years & 1,2,3 \\
P5  & F  & Master & 4 years & 1,2,3 \\
P6  & M  & Doctorate & 18 years & 1,2,3 \\
P7  & M  & Master & 2 years & 1,2,3 \\
P8  & M  & Master & 8 years & 1,2,3 \\
P9  & M  & Master & 1 year & 1,2,3 \\
P10 & F  & Doctorate & 30 years & 1,2,3 \\
P11 & F  & Master & 2 years & 1,2,3 \\
P12 & M  & Doctorate & 7 years & 1,2,3 \\
P13 & M  & Doctorate & 5 years & 1,2,3 \\ 
P14 & F  & Bachelor & 2.5 years & 3 \\ 
P15 & F  & Bachelor & 2.5 & 3 \\ 
P16 & M  & Doctorate & 3 years & 3 \\ 
P17 & M  & Doctorate & 6 years & 3 \\ 
P18 & M  & Master & 2 years & 3 \\
P19 & F & Doctorate & 2 years & 3 \\ 
P20 & M & Master & 2.5 years & 3\\
P21 & M & Bachelor & 3 years & 3\\ 
\hline\hline
\end{tabular}%

\label{tab:demo}
\end{table}

\subsection{Results}
In this section, we present our findings from the interviews, highlighting the various challenging sensemaking scenarios experts encounter in real-world deployments, along with needs and opportunities for assistance.

\subsubsection{Challenging Sensemaking Scenarios in Current Practices}
The first significant challenging scenario is contextualizing the data given the highly personalized nature of an individual's life.
The experts we interviewed mentioned encountering people living in unconventional settings such as boats or RVs, or with conditions like using a cane, making it difficult to create a unified approach to interpret their behaviors.
However, in many cases, researchers do not leverage critical information about a person's lifestyle during data analysis. 
Our participants highlighted a concerning issue where \emph{``none of that (user information) gets posted to the data afterward''} and is \emph{``lost track of''}.

Another challenge stems from unpredictable human decisions, which create a vast array of possible events. 
P6 recounted an incident where a participant accidentally dislodged a motion sensor and stored it in a drawer without notifying their research team. 
It took a while for the team to notice the \emph{``little spurts every so often''}, and they could not deduce what had happened until talking with their participants. 
Due to the inherent limitations of passive sensing systems, experts often have to speculate about the events behind the data, which requires them to rely on their own experiences and imagination, which could be restrictive and even introduce biases in the inference.
Thus, P9 expressed a desire for creative interpretations that could offer insights beyond their initial understanding as useful inspiration.

The example of the fallen motion sensor reveals a prevalent scenario: capturing and explaining anomalies. 
Nearly all experts point out that things can easily go awry in a longitudinal study, where identifying the cause of an anomaly is challenging due to the lack of a straightforward correlation between issues and their causes.
There are inherent data limitations that experts must consider, for example, GPS location from the phone is only valid if the participant was carrying their phone with them (P3). 
Successfully identifying and rationalizing anomalies demands a deep understanding of both system limitations and human behavior, a skill that typically only experts possess.
However, even for experts, the task remains cognitively challenging as it often requires cross-referencing multiple data sources to determine whether an anomaly truly indicates a problem and, if so, why it occurred. 

Another way to obtain an interpretation of data is to source directly from participants. 
However, discussions with experts reveal another significant challenge: understanding and utilizing self-report data. 
Self-report data is often limited, and experts must be cautious in treating self-report data as absolute truth, especially when dealing with older adults or those with cognitive impairments (P4). 
Even when experts manage to obtain some level of ground truth, they still face the challenge of integrating this information with other data modalities, mirroring the process of contextualizing raw sensor data.
This gap makes it difficult for experts to generate and validate their interpretations.

Challenges in obtaining self-report highlight another crucial but difficult scenario: determining the necessary actions after data analysis, often involving multiple stakeholders.  
A common process observed involves experts developing insights and then communicating these with various stakeholders to ensure that interpretations of sensing data accurately reflect real events. 
However, key stakeholders that can help the interpretation of data, such as participants themselves or their caregivers, are not always readily accessible to researchers.
As a result, experts described a cautious process in which they often perform internal checks to make sense of the data before reaching out, ensuring that any contact with stakeholders is necessary and meaningful. 
This further underscores the necessity of an initial sensemaking process by experts in many real-world deployment scenarios to reduce the burden on both researchers and participants by minimizing the need for frequent check-ins.

Based on our interviews, we summarize a list of \textbf{scenarios} for the system to assist with experts' sensemaking: 1) \textbf{contextualizing} data based on \textbf{personalized schedules} and \textbf{self-report data}, 2) generating insights on various \textbf{possible events}, 3) capturing and explaining \textbf{anomalies}, and 4) determining \textbf{actions}) for interacting with different stakeholders. 
Next, we discuss the needs and opportunities of assisting researchers' sensemaking.

\seclabel{gap}

\subsubsection{Needs and Opportunities in Assisting Experts} 
A key insight emerged from our interviews: experts' sensemaking is inherently \textit{iterative}.
P6 mentioned that in real-world practice while making sense of data, they are \emph{``constantly developing and improving.''} that oscillates between \emph{``what we want to measure''} and \emph{``what is available for us.''} 
To better understand this iterative process, we identified two distinct approaches: one starting from the data and moving towards the insights, and the other beginning with the goals and working towards the data.

The first approach to examine the sensing data to identify any emerging patterns or insights allows experts to uncover unexpected trends and behaviors that might not have been initially considered. 
For instance, P8 remarked, \emph{``it wasn't until we got into the project that we realized maybe there are other aspects we need to be looking at.'' } 
P6 likened the process to a \emph{``fishing expedition,''}, explaining that 
\emph{``more people are always coming up with new and interesting ways of extrapolating info out of this data.''(P6)}. 
Starting with the data enables experts to discover novel, previously undocumented insights and generate ideas that could lead to a deeper understanding of the data or behaviors of the observed participants.

The second approach prioritizes identifying goals such as activities of interest from the outset. 
Experts initially concentrate on specific activities or behaviors that are crucial for their study or intervention. 
They then examine the available sensors and data to ascertain how to obtain relevant information about the identified activities. 
This method is more targeted, starting with a precise idea of what needs to be observed and then devising the best strategies to collect and interpret the required data. 
As P12 noted, \emph{``which activities you look at depends on your research goal'' }. 
Although these two paths differ, both are integral in the sensemaking process. 
Almost all experts confirmed that they frequently use these approaches in tandem to enhance their understanding and adjust their investigative focus as new insights emerge. 
P12 highlighted the complementary nature of these approaches, stating \emph{``both approaches have their place.''} 

Now, recognizing the iterative nature of the sensemaking process and the two distinct paths it can take, we must consider how we can support this process effectively. 
We have identified two primary needs based on these paths: \textbf{direct representation} of data and \textbf{indirect inference}.
Experts require a direct representation of sensing data to manage the vast volumes of information, as sensor data often does not come in a human-readable format. 
Throughout our interviews, experts commonly relied on visualizations for this purpose. However, most visualizations lack systematic design, are overly generic, and are often created ad hoc for specific tasks. 
Experts mentioned techniques like stacking modalities or creating correlation graphs, but these efforts are sporadic and have not been standardized even within the same research group.
Beyond direct data representation, participants also expressed a need for assistance with automatic inference generation as potential guidance for their sensemaking.
Even with effective visualizations, the transformation of data into actionable insights is still predominantly a manual process that the experts conduct\jiachen{, which is overwhelming when dealing with so much data (P9).}
In longitudinal studies that span months or even years, it becomes impractical for experts to examine every piece of data to get the desired granularity required for some applications like labels for machine learning (P12/13).
As a result, assistance in quickly navigating to the desired information seemed crucial.
Our participants noted a significant lack of tools for this purpose, as the sensemaking process is complex and often beyond the capabilities of current algorithms and tools. 
For instance, some experts who used machine learning in their studies mentioned that they only viewed benchmarks as ``results to be compared'' against ground truth data, rather than as useful assistance for them. 
However, they expressed interest in getting some extracted data insights that go beyond simple labels.
Based on these insights, we designed and developed our initial prototype. 

\section{INITIAL PROTOTYPE: CONNECT DATA WITH INSIGHTS} 
\label{sec:system}
In this section, we present the proposed design of our initial prototype, \textbf{Vital Insight version 1 (VI-1)}, based on the four sensemaking scenarios and two primary goals. 
The design of VI-1 aimed to reflect the insights from the preliminary interviews as a prototype \jiachen{and underwent several iterations through casual group discussions with experts}.
VI-1 will be used as a technology probe to assist experts in different sensemaking tasks to generate high-level insights~\cite{hutchinson2003technology}.

\subsection{Data Types and System Infrastructure}
The input for VI-1 could be a variety of data types commonly used in real-world passive sensing deployments from smartphones, wearables, and voice assistants/chatbots. 
These data types include time-series data (e.g., physiological signals and mobility patterns), discrete data (e.g., phone unlock/lock states), and self-reported data (e.g., survey responses, daily check-in conversation with chatbots) -- all commonly used data elements in ubiquitous computing studies that involve passive sensing~\cite{Fritz2014PersuasiveTI,Loureiro2012SensingTA}.
VI-1 provides a combination of visual data representations and text-based descriptions. 
We built the system using React, with a MongoDB backend hosted on a local server to ensure fast data retrieval and robust security protections.
 \begin{figure}[htbp]
  \centering
  \includegraphics[width=\linewidth]{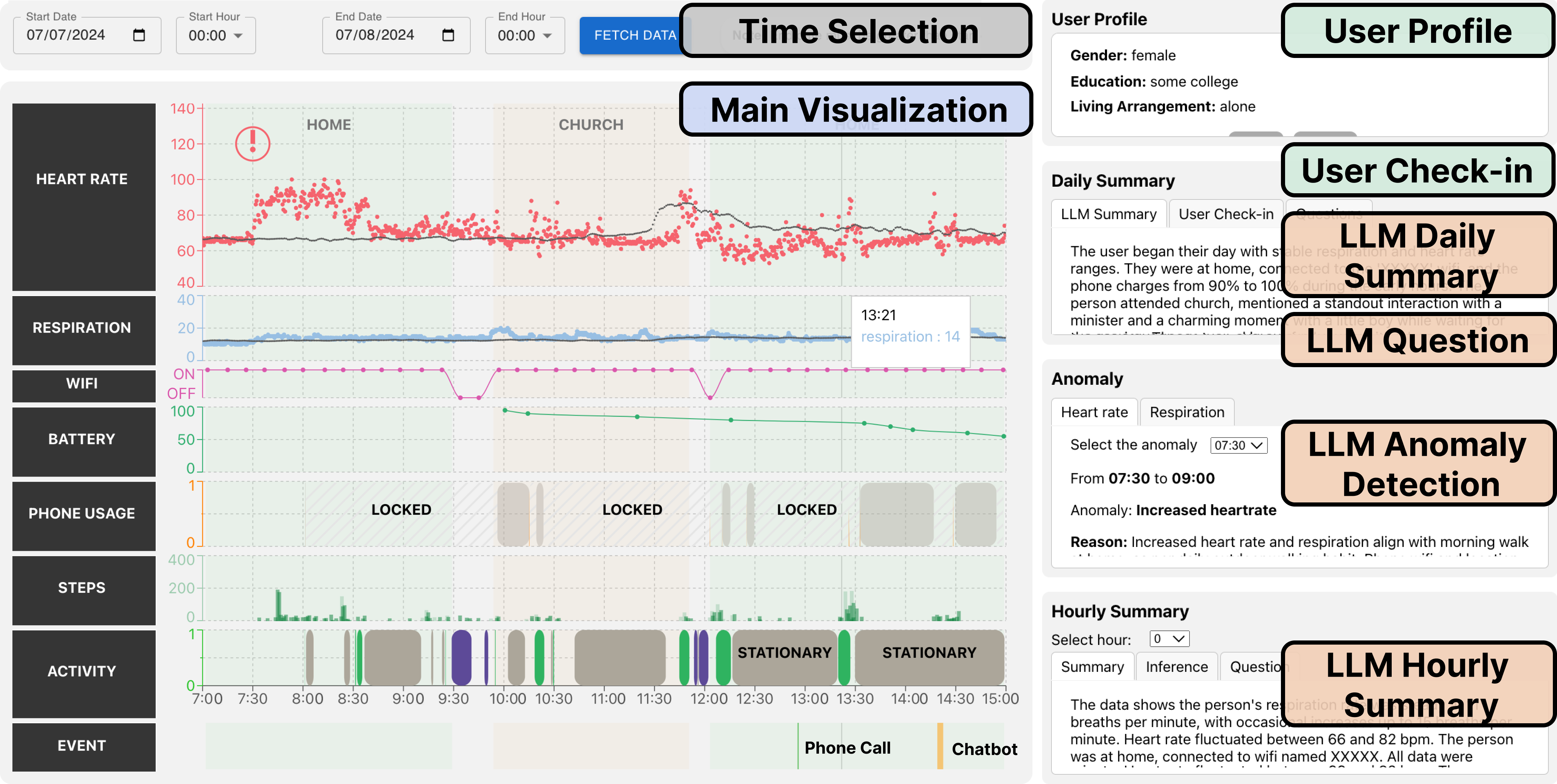}
  \caption{Eight components in VI-1 powered by visualization and LLM: time selection, main visualization, user profile, user check-in (user conversation with chatbot), LLM daily summary, LLM question, LLM anomaly detection, LLM hourly summary}
  \label{fig:vi1}
  \Description{Screenshot of the Vital Insight interface showing the User Profile on the left with details like gender, education, race, marital status, occupation, living arrangement, addresses, and recreation activities. On the right, two Anomaly panels display heart rate and respiration issues, with timestamps and explanations for spikes in heart rate and increased respiration related to Rapid Eye Movement sleep.}
\end{figure}

\subsection{Interface Design}
Our initial prototype has two major parts: (1) visualizations that provide direct representation of sensor data, and (2) varying granularities of summaries and inferences generated by LLM that provide indirect inference. 
Our system interface has five panels with eight components: time selection, main visualization, user profile, daily summary (user check-in, LLM-generated daily summary and questions), LLM-anomaly, and LLM-hourly summary (Fig. \ref{fig:vi1}).
To generate the interface, our system uses data from a real study with participants, which we discuss in greater detail in Sec. 4.3.

The \textbf{Time Selection} panel (top left) allows users to select the start and end date/time to zoom in/out, and iteratively switch between a holistic view and details on demand \jiachen{to facilitate customization}, in line with best practices for visual information seeking through data visualization~\cite{shneiderman2003eyes,wootton2024charting}.
The \textbf{User Profile} panel on the top right consists of basic demographic details such as age and living situation, designed to provide experts with a holistic understanding of the participant's lifestyle \jiachen{(Scenario 1)}.

The \textbf{Main Visualization} panel serves as the central component to represent the sensing data (Fig. ~\ref{fig:vi1} left). 
The visualization component features five categories of information gathered from the phone, smartwatch, and chatbot: location, health, phone usage, activity, and events. We used a small multiple-style time series plot arrangement to display each category of information along a unified timeline. 
\jiachen{The design also aligns with current practices, where experts manually stack different modalities of data, as described in the interview (Section 3.2.2).}
The \textbf{Location} in the background shows location labels, e.g., home and church. We generate this data by cross-referencing the raw GPS coordinates from the phone app with participants' self-reported frequently visited locations and addresses. The system marks each location with a unique background color across all views with a label at the top, aiming to provide contextual information when experts examine the data.
In Fig. \ref{fig:vi1}, the plotted location labels are home (light green) and church (light orange).
The \textbf{Health} view presents heart rate and respiration data from the smartwatch using scatter plots in red and blue.
Two black trend lines overlayed on the graph represent the participant's average heart rate and respiration over a one-month period. Exclamation marks (!) next to the scatter plot indicate detected anomalies. 
We present more information about the anomaly detection module in section 4.3.
The \textbf{Phone Usage} view includes Wi-Fi (connected/disconnected, purple line chart), phone battery levels (0-100, green line chart), and phone interaction (unlock/lock events, solid/dashed grey rectangle overlaid). 
The \textbf{Activity} view reports step counts and basic activity recognition. 
A bar chart presents the step counts, with the height of the bar representing the number of steps taken during the time indicated by the width. 
The activity view presents basic activities from iPhone's standard activity recognition API, using different colored rectangles (‘stationary’ in gray, ’walking’ in green, ’automotive’ in purple), indicating the activities and their respective time spans.
Finally, the \textbf{Event} view includes significant events throughout the day, including phone calls (green) and interactions with the Alexa chatbot (orange), and their respective durations.


\textbf{Daily Summary} panel, provides a high-level overview of the participant's day with three tabs: LLM Summary, User Check-in, and Questions. 
The \textbf{LLM Summary} tab presents an LLM-generated summary of the entire day in paragraphs and bullet points for easy reading, thus providing experts the ability to quickly understand a person's day. 
\jiachen{We selected LLM to generate the summary due to its extensive capability to customize text-format summaries and explanations, offering different levels of detail and explainability for experts.}
The \textbf{User Check-in} tab displays the real conversation between the chatbot and the participant (orange `Chatbot' event in visualization). 
The \textbf{Questions} tab presents a list of questions and missing information that the system identifies as potentially helpful for data interpretation. 
This information can serve as an inspiration for experts to reflect on potential actions, particularly when engaging with different stakeholders, such as speaking directly with participants \jiachen{(Scenario 4)}.
We present detailed rationales for the LLM-generated content in Sec. 4.3.

The \textbf{Anomaly} panel provides insights about \jiachen{anomalies and outliers, typically defined as ``an observation that deviates significantly from other observations, raising suspicions that it was generated by a different mechanism,'' and is crucial in analyzing time-series data such as heart rate~\cite{hawkins1980identification}. 
In previous interviews, experts noted that anomalies are worth paying attention to (Scenario 3).}
For each anomaly, the panel displays the specific time range, description, and a possible reason, all generated by the LLM. 
Users can select the specific anomaly time to view the detailed information.
The \textbf{Hourly Summary} panel provides LLM-generated interpretations for each hour of data. 
Users can select a specific time (e.g. 4-5 AM) to view the corresponding interpretation. 
Similar to the daily summary, this panel offers an objective summary of the sensor data with minimal inference, more creative insights derived from the data, and questions that the LLM identifies as useful for interpreting that hour's data. 
This panel offers different levels of inferences on potential events in detail \jiachen{(Scenario 2)}.


Additionally, we implemented several interactive functions to improve the system's usability. 
To enhance the visual alignment of different modalities, we added a vertical tooltip that appears across all modalities when the user hovers over the visualization and displays the time and value.
This helps experts easily compare and integrate different sensing modalities for interpretation.
To avoid label overlap on a sequence of narrow rectangular boxes of phone unlock/lock and activity, we implemented a hover effect that reveals labels (e.g., ``walking'') only when a user hovers over them.


\subsection{LLM Augmentation}
Our initial prototype provides four main LLM functionalities: the hourly summary, daily summary, question, and anomaly, all developed based on the OpenAI GPT-4\jiachen{o-mini} model through zero-shot Natural Language String method \jiachen{using static prompts} (Fig.~\ref{fig:llm})~\cite{kim2024health,gruver2024large,liu2023large}. 
We provide detailed prompts in the Supplementary Materials.

There are two primary forms of data input: Sensing Data and Contextual Data.
For \textbf{Sensing Data}, we process the data into \textbf{semantic description} \jiachen{before sending it to the model}. 
Discrete data, such as WiFi and phone battery levels, are directly converted into sentences with values and timestamps (\emph{e.g. ``The battery level of the person's phone is [BATTERY] at [TIME].''}). 
For time-series data like heart rate, encoding each data point individually would overwhelm our system. 
Instead, since these data are typically sampled at a fixed rate, we group them and send them in an array format like \emph{``The person's respiration from [TIME] to [TIME] (10-second interval) is [VALUE1, VALUE2, \ldots].''}, and sort this narrative in chronological order.
Another data input is \textbf{Contextual data}, which includes User Profile - demographic and routine information, and User Check-in - where we process conversation in a time\emph{(From [TIME] to [TIME])} and utterance \emph{([ROLE]:[UTTERANCE])} format.
\jiachen{We do not specifically handle missing or noisy data and transfer each data point as-is.}

 \begin{figure}[htbp]
  \centering
  \includegraphics[width=\linewidth]{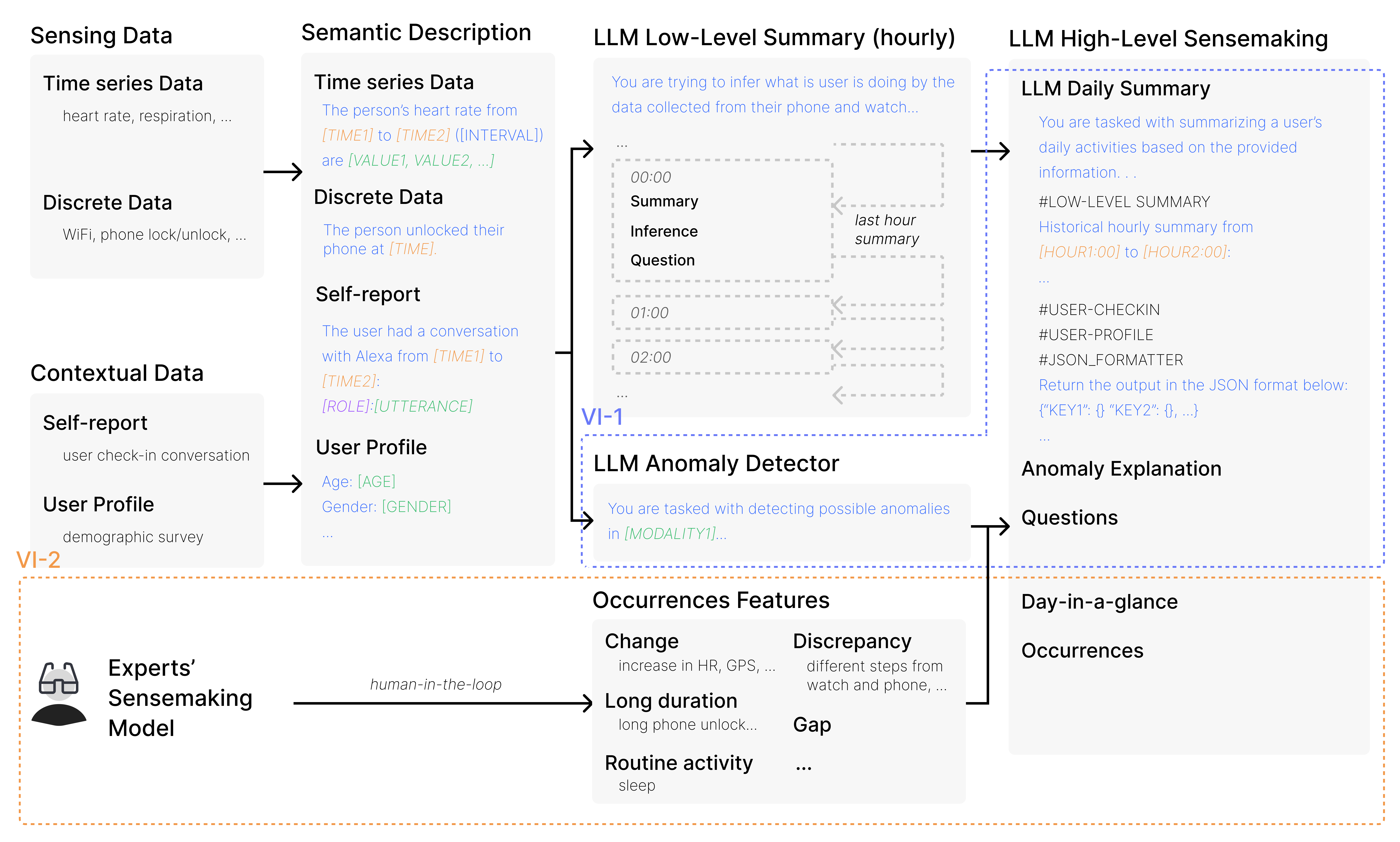}
  \caption{Overview of the structure of LLM augmentation for VI-1 and VI-2.}
  \label{fig:llm}
  \Description{Diagram showing the structure of the LLM system with two main sections: Input and Output in Vital Insight. The Input section includes 'Sensor Data' labeled as Semantic Sensing Data and 'Contextual Information' with User Profile, User Check-in, and Daily Feature. Arrows connect these inputs to the Output section, which features 'Hourly Data' with tabs for Summary, Inference, and Question. Outputs from Hourly Data lead to 'Daily Summary' for LLM Summary, 'Question' for high-level questions, and 'Anomaly' for detection and explanation.}
\end{figure}

After transferring raw data into semantic descriptions, \jiachen{we send these descriptions to the model, which } generates \textbf{LLM low-level summary}, processing the most granular level of data and providing an hourly summary. 
The prompts consist of the following components: Goal, Data Interpretation Guidance, Data, User Profile, User Check-in, Historical Summary from GPT, and a JSON Formatter. 
The Goal outlines the task's objective and explains the purpose of each input component. 
Data Interpretation Guidance offers expert advice on analyzing the data.
Data includes one hour of sensing data in the semantic format along with the User Profile and User Check-in. 
The Historical Summary from GPT retrieves outputs from the previous hour's LLM-generated summary as additional contextual information. 
The JSON Formatter specifies the desired JSON output format. 
It outputs summary, inference, and questions as described in Sec. 4.2. 
 
The prototype then processes the low-level summary into \textbf{LLM High-Level sensemaking}. 
Recognizing the daily patterns in activity, it aggregates the outputs from 24 hours of the hourly summary and other contextual information to generate a similar output containing a summary, inference, and questions, but on a daily basis. 
The prompt includes the Goal (\emph{``You are tasked with summarizing a user's daily activities based on the provided information\ldots''}), User Profile, User Check-in, Historical Summary from GPT (24-hour output from the hourly function), and the JSON Formatter. 
The system generates and formats the results in a human-readable way, both as a paragraph and in bullet points for easier digestion.
It also summarizes information like questions at a higher daily level over a 24-hour period.
\jiachen{Before sending the summary to the dashboard, we perform several programmatic checks and rerun the process if the output does not meet the standard (e.g., output tokens<10). }

Other than summaries from the hourly data, \textbf{LLM anomaly detector} identifies potential anomalies in different modalities directly using the semantic descriptions of each sensing modality.
While we iteratively adjusted these prompts to optimize outputs, we provided all LLM-generated results as-is and unaltered during user testing.
\jiachen{If the inputs exceed token limits, we split them into chunks before sending them to the model; however, after this multi-level procedure, it's very unlikely that the input tokens will exceed the context window of the model (128k).}

Past research has used similar methods to embed sensing data into narratives before sending to LLMs~\cite{ji2024hargpt,cosentino2024towards,yang2024drhouse,kim2024health,ouyang2024llmsense,fang2024physiollm}. 
While not the main focus of this paper, our method specifically distinguishes between different types of sensing data and identifies appropriate ways to handle each type.
We also developed a multi-level pipeline that integrates contextual information at each step of the inference process, generating different inferences at various levels of granularity. \jiachen{This multi-level pipeline also avoids sending an extensive amount of data to the model, which has been proved to be problematic and has led to inaccurate LLM outputs~\cite{levy2024same}.}
At the same time, we realized that embedding sensing data into LLMs is an active and rapidly growing research area. Therefore, we designed VI-1 to keep the backend LLM and the frontend display of information relatively separate to ensure easy integration of advanced embedding methods into VI in the future.


\subsection{Alignment with Interview Insights}
We designed VI-1 based on insights from the initial interviews. 
In this section, we briefly describe how the features align with those insights.
From the interviews, we identified four key sensemaking scenarios that guided the feature design. 
To address the scenario of \textbf{contextualizing data based on personalized schedules and self-reported data}, we incorporated user check-in and profile modules to provide direct information about the person. 
Additionally, important contexts such as location labels are integrated into the visualization to aid experts in contextualizing the data. 
The LLM also incorporates various contextual information to enhance its inferences.
To \textbf{generate insights on various possible events}, the LLM provides daily and hourly summaries to help experts understand events at different levels of granularity. 
We designed the anomaly panels to specifically address the scenario of \textbf{capturing and explaining anomalies} with a quick glance.
Lastly, the question panel offers inspiration for \textbf{actions and interactions with different stakeholders}, providing valuable guidance for experts on specific aspects of the data to inspect further.
Additionally, the design of VI-1, incorporating both visualization and LLM inference, aims to assist with both \textbf{direct representation} and \textbf{indirect inference}.

\section{STUDY 2: EXPLORATORY SESSION} 
\seclabel{evaluation}
After developing the first prototype, VI-1, we conducted user testing with 13 experts to explore the potential of VI-1 in assisting experts in different sensemaking tasks. 
Through these interactions, we also used VI-1 as a technology probe to observe experts' sensemaking process in integrating multiple modalities of information. 
In this section, we detail the protocol of the user testing sessions, along with the qualitative and descriptive findings from the sessions.

\subsection{Method}
\label{sec:method_testing}

\subsubsection{Data Collection}
To generate the dashboard used in the user testing session, we used real-world data from a deployment with participants, where we collected: 1) pre-study surveys, 2) passive mobile and wearable sensing data, and 3) voice assistant check-ins.
Participants completed surveys on demographics and regular routines before deployment.
We provided the participants with a Garmin smartwatch and installed the study app on their phones. 
Our app directly syncs with the Garmin device and collects their physical activity, location, call logs, app usage, Wi-Fi/Bluetooth connections, IBI, heart rate, accelerometer, step counts and more.
Additionally, participants can initiate a chit-chat conversation with an LLM-powered chatbot on the study provided Amazon Alexa to check in on their day.
The system has been deployed in prior studies~\cite{campbell2023patient,mishra2021detecting,kunzler2019exploring,mishra2022towards,aip_2024}, and runs in the background, continuously storing and uploading data to our secure server. 
We used a secure, HIPAA-compliant GPT-4 model that does not store data for retraining. Additionally, we ensured that no identifiable data is transmitted to the model.
For example, we only send the de-identified location labels such as `home' to the external API services instead of raw GPS location, and manually remove potentially identifiable Wi-Fi names and user check-ins.
The dashboard presented to the experts does not contain any identifiable information. 
We used tokens to provide temporary access to the system during the user testing.
These precautions ensure that neither GPT-4 nor the participants can identify the subject of data collection.
The individuals who provided data consented to share it with external researchers through an IRB-approved study.


\subsubsection{Study Design}
At the beginning of the session, we provided experts with a brief tutorial on VI-1.
We then gave them a link to the system and asked them to share their screen as they completed a series of tasks with the information on the system. 
As an exploratory session, we used the same day of data for all the user testing sessions for consistency.
We designed these tasks to specifically address the four sensemaking scenarios we identified during the initial interviews.
For the first task, we asked experts to use all available information on the system and write a short, general summary of the person's day. 
We kept the task broad, allowing experts to determine their own version of what the person's day entailed.
Next, experts were asked to dive deeper into specific activities throughout the day in greater detail.
We then focused on anomalies, asked experts to identify potential anomalies in the data, and guided them to the corresponding panels to share their thoughts on the results from the system.
Finally, we discussed interaction with stakeholders, directing experts to the question section and discussing possible actions they might take based on the information provided.
After completing these tasks, we had an open-ended discussion to gather their general thoughts on the system and provided them with a survey.
In the survey, we first asked experts to evaluate the different modules of the system. 
They rated each module based on its importance, trustworthiness, utilization, and clarity\jiachen{, common metrics in evaluation visualization based dashboard~\cite{10360907,almasi2023usability}}. 


Similar to the initial interviews, we conducted hour-long user testing sessions online with 13 experts and gave a \$15 gift card as compensation. 
We recorded all user testing sessions, capturing the experts’ interactions with the prototype.
For the qualitative analysis, we obtained transcripts of the sessions and the summaries that experts wrote during the first task. 
We also took observation notes during each user testing session. 
We then followed an open-coding approach to analyze all the materials, the same method as our initial interview study (Sec. 3.1). 
The analysis focused on two main themes: the general usability of the prototype and the sensemaking process of different experts.
We also conducted a quantitative analysis of the survey results.

\begin{figure}[htbp]
  \centering
  \includegraphics[width=\linewidth]{figs/des_s2.png}
  \caption{Survey results from the initial user testing (Study 2).}
  \label{fig:des_s2}
  \Description{}
\end{figure}

\subsection{Prototype Evaluation Results}
In this section, we first present the quantitative results from the survey, followed by an explanation of general feedback on the prototype based on the qualitative interviews. 

\subsubsection{Survey Results}
First, we report the results of the survey.
Regarding the \textbf{impact} of different modules on sensemaking.
The results indicate that Visualization was clearly the most crucial component, while Hourly Summary was the least, with the other three modules (User Check-in, Daily Summary, and Anomaly) being similarly important.
In terms of \textbf{trust} in different modules (scored 1=Distrust to 5=Trust), experts trusted User Check-ins (Mean 4.37, SD=0.47) and Visualization (4.32, SD=0.50) the most, followed by Daily Summary (3.91, SD=0.75), Hourly Summary (3.73, SD=0.69), and Anomaly (3.69, SD=1.03) (Fig.~\ref{fig:des_s2}). 
The average trustworthiness score of the system as a whole was 4.09 (SD=0.70). This suggests that the experts generally considered the system to be trustworthy, with the indirect inference components by LLM rated slightly less than the direct data representations. 
The lower trust in the Anomaly module may also be due to differing definitions of what constitutes an anomaly and vague guidance on how it was determined.
Next, we examined the \textbf{usage} and \textbf{clarity} of each data view in the visualization. 
Among all seven data views in the visualization, experts used an average number of 6.23 (SD=1.01). Heart rate, steps, and activity were used by all experts (13), followed by respiration (12), WiFi (11), battery (10), and phone unlock usage (9).
Regarding the clarity of each modality (scored 1 = Very unclear to 5 = Very clear), the average score across all seven modalities was 4.29 (SD=0.37). The clarity ratings from most to least clear were battery (4.77, SD=0.44), WiFi (4.62, SD=0.51), phone unlock usage (4.54, SD=0.52), heart rate (4.23, SD=1.3), respiration (4.08, SD=1.26), steps (4, SD=1.15), and activity (3.77, SD=1.17). 
These findings show that most data views were generally clear, and experts actively used this information during their sensemaking process\jiachen{, which aligns with other studies on visualization systems that also consider a similar score to be a good rating~\cite{10.1145/3613904.3642237,10.1145/3613904.3643022,zhu2007measuring,north2006toward}}.

\subsubsection{General Usability and Qualitative Feedback}
Experts generally liked the prototype, with P4 expressing, \emph{``I wish I had this for my study.''} 
While we covered and evaluated many aspects of the prototype in the surveys, we wanted to explore the qualitative feedback in greater detail to provide more context, and identify aspects not captured in the survey.

Experts loved the visualization aspect of the prototype. 
For example, P3 specifically mentioned that they appreciated how the steps from both the phone and watch were plotted together. 
Further, experts found the User Profile module helpful (\emph{``The user profile, that's very powerful(P1)''}). 
Similarly, experts loved the user check-in module and wanted it to be more prominent on the dashboard.
Some experts like P1 suggested \textbf{extracting key information from the user check-in} panel and displaying it on the visualization according to the time. 
For example, P1 suggested it would be helpful \emph{``if the dashboard said, Oh, this is when the little boy sat. This is when the brunch happened.''} as an addition to the original paragraph. 
Additionally, P4 and others suggested displaying timestamps along with some semantics for the conversation with the chatbot in the user check-in module, other than only as an event on the visualization.

During the user testing sessions, many experts appreciated the rich details in both the visualizations and LLM-generated summaries. 
Many experts also liked the bullet points in the summaries, noting that they provided a good overview.
However, experts suggested that it could be even more effective if there were an additional high-level summary beyond the bullet points, such as \emph{“more highlights of the day”} (P1). 
This may also be due to the limited time given for experts to complete the tasks, requiring them to explore the system quickly and make assumptions. 
This feedback indicates that even experts who focus on detailed information still value a simple summary, like \emph{``this is a normal Sunday''} as a starting point.


Many experts found the LLM-generated summaries to be \emph{``good''} and \emph{``beyond expectations.''} 
Experts generally agreed that the system is evidence-based, which aligns with the survey results. 
Through interaction with the various components, many experts were able to identify the evidence used by the LLM to make its assumptions.
Experts also appreciated the different possibilities of events provided by the LLM, and enjoyed the current ways in which evidence is generally separated from inference. 
However, many expressed a desire for the evidence to be more apparent, such as by highlighting the data used in the visualization or indicating when information is derived from user check-ins. 
P5 mentioned:
\emph{``I don't mind making some guess, but I wanted to know where that info (from LLM) is coming from.''}

Another interesting insight emerged when examining the anomaly: the differing \textbf{definitions of} what should be considered an \textbf{anomaly}. 
Many experts noted that the current anomalies are often more like \emph{``standout events''} rather than true anomalies or alerts.
Almost all experts confirmed that there were hardly any anomalies needing immediate concern in the current data, and emphasized they would need to better understand the person's regular patterns and routines to further classify anomalies. 
This highlights a potential need for the prototype to distinguish more clearly between special events or highlights and genuine anomalies that experts should be concerned about, which we will discuss further in the next section.

For the Hourly Summary and Questions, some experts actively used them, but similar to the survey results, many experts found these modules to be \emph{``a little bit too much (hourly)''} or \emph{``confusing at the beginning (question).''} 
Experts did confirm their accuracy regardless.
For example, one expert with over 10 years of experience in this field noted, \emph{``These are very similar questions that I would ask them.''}
We realized that while this information may be helpful, it is not the most direct form of assistance for experts' sensemaking process.


In summary, the qualitative insights from experts generally aligned with the quantitative survey results, indicating that experts had a positive experience using the prototype. 
Their feedback on the prototype included simplifying the daily summary, adding/extracting more details for the user check-in, and clarifying the definition of anomaly. 
In the next section, we will examine experts' sensemaking process in detail, exploring how they navigate between different data views to generate insights.
These insights can further inform modifications to the prototype, enabling it to better support experts in their sensemaking process.

\subsubsection{User Interaction: Experts' Sensemaking Process}
In this section, we further examine experts' thought processes in sensemaking through their self-descriptions (think-aloud) and recorded screen interactions.
Experts employ a variety of approaches for sensemaking, with notable commonalities as well as differences. 
Most experts started their analysis with the visualization (P4/5/6/8/10/11/12/13), while P9 started with the LLM summary, and P1/2/3/7 began with the user profile. 
Experts quickly switched to the other to verify details, moving back and forth between the visualization and the LLM contents.
Some experts composed the summary entirely on their own, whereas others opted to copy and paste the LLM inferences into the summary and edit (P8, P13), such as removing information they deemed trivial or unimportant (e.g., WiFi details, specific phone usage patterns) and adding significant details not mentioned in the LLM-generated summary (e.g., sleep patterns, correlations between activity, and heart rate).

To further scaffold their sensemaking process, we first confirmed that experts' sensemaking process is inherently \textbf{iterative}, moving from high-level insights to detailed analysis and then back to a broader perspective. 
Through the iterative process, experts \textbf{integrate multiple modalities} to make assumptions. 
For closely linked modalities, some processes also served as a form of validation. 
For instance, P11 initially assumed that a decrease in battery level indicated screen time, but rejected this assumption on noticing that the phone was locked during that period. 
Validation across modalities was frequent and varied, with common comparisons between WiFi, location, heart rate, respiration, steps, and activity. 
Experts also attempted to integrate and compare user check-ins with other sensing modalities. However, this was more difficult due to the current setup, where user check-ins are fully separate from the main visualization. 
As a result, they expressed the need for the LLM to extract check-in information and display it on the visualization to facilitate easier comparison between modalities.

An interesting variation in sensemaking among experts is whether they proceed in \textbf{chronological} order. 
There were notable deviations from this method, revealing different strategies that were not bound by time sequence.
One common non-chronological approach involved starting with \textbf{noticeable patterns}. 
In the previous section, we found that experts defined anomalies differently compared to the LLM. At the same time, they confirmed that while anomalies were important, they were relatively rare, and other patterns were also worth noting.
So, what are these ``standout events'' that experts identified?
We summarized six patterns.
Experts often noticed \textbf{changes}, and \textbf{gaps} in the data. 
For instance, P6 noticed and examined the change to `automotive' from a long time of   `stationary', and many experts investigated and even asked questions about the missing activity records and location labels at certain times. 
The \textbf{long duration} of certain events is also crucial, such as extended phone usage or prolonged sedentary periods.
In several instances, experts identified \textbf{discrepancies}, such as P6 noting the presence of steps while the subject appeared to be stationary or missing activity data.
Another non-chronological sensemaking approach involved identifying key \textbf{routine activities}. 
For instance, P4/6/8 instinctively prioritized identifying sleep without being explicitly tasked to do so, later confirming in discussions that sleep is a critical activity they knew was important based on their own experience.

We also observed some interesting interactions with the LLM-generated inferences. Experts generally displayed low initial trust in these inferences, often \textbf{questioning and verifying} the results rather than accepting them as-is: 
\emph{``I was trusting enough to go and try to see it\ldots\ but I still want to verify it as well.'' (P9)}
A common scenario where experts questioned the inferences was when they read the LLM-generated summary before reviewing user check-in data and could not immediately \textbf{identify the source of the information}. 
For example, P6 questioned the summary: \emph{``Evening afternoon nap\ldots How was that detected, huh?'' }
They later found the relevant information in the user check-in, which clarified their confusion. 
As experts verified the inferences and found that the information aligned with the raw data, their \textbf{trust} in the LLM-generated summaries grew. 
P13 shared this sentiment when checking an LLM-generated anomaly explanation about step counts around 2 AM: \emph{``Then I noticed, like, the steps actually have some number value above 0 (on the visualization). So actually, that makes sense.''} 

Most experts, given their expertise in data interpretation, preferred to form their \textbf{own judgments first} to avoid bias before comparing with LLM results. 
P2 explained: 
\begin{quote}
\emph{``I was hesitating to check out the summary to be the 1st thing because I would be worried of being biased.''}
\end{quote}
Still, experts found the LLM summaries valuable, especially as a guide for further exploration. 
For instance, P9 skimmed through the LLM summary before diving into the visualization details to \emph{``get a feel of what it saw first or what I was able to summarize.'' }
They noted this approach was particularly useful when they were \textbf{unfamiliar} with the participant: 
Other experts echoed that the LLM summaries helped them \textbf{notice} aspects they might have otherwise \textbf{overlooked} due to the sheer volume of data. 
P7 remarked on a second anomaly: 
\begin{quote}
\emph{``If not reading what was generated by the LLM, I didn't notice there's a small increase in steps.''}
\end{quote}

In addition to the LLM summary, we observed common usage patterns for other panels when experts were making sense of the data. 
The user profile helped experts familiarize themselves with the participants and allowed them to associate possible routine activities with data. 
The user check-in served as contextual information, confirming activities like going to church and \textbf{filling in details} not available in the visualization, such as afternoon naps, activities during church, and brunch with friends. 
All these insights highlight the varied usage of different modules among experts.

\begin{figure}[htbp]
  \centering
  \includegraphics[width=\linewidth]{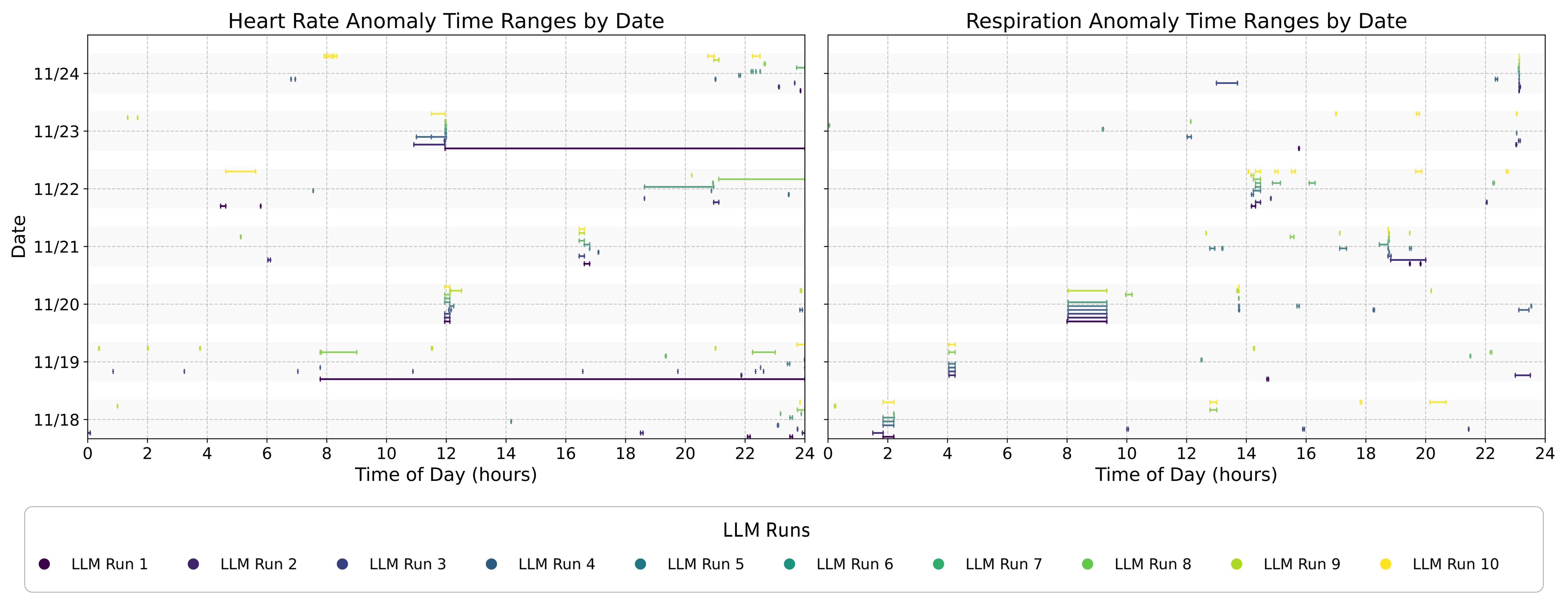}
  \caption{\jiachen{Time ranges of LLM detected heart rate and respiration anomalies from 11/18/2024 to 11/24/2024. The horizontal lines represent the start and end time of the anomalies, and the different colors represent 10 LLM test runs.}}
  \label{fig:anomaly.png}
  \Description{}
\end{figure}
\subsection{Follow-up Evaluation on LLM Anomaly Detection}
\jiachen{Through the interviews, we discovered that the LLM's anomaly detection on a single modality did not align with the experts' expectations. As a result, we conducted an exploratory analysis to evaluate LLM anomaly detection. We ran the LLM anomaly detection on heart rate and respiration data for a week (11/18/2024 - 11/24/2024), performing 10 tests for each day using the same prompts. We observed significant instability in the detection on several days (Figure. ~\ref{fig:anomaly.png}). Ideally, all the different colors of the horizontal lines should align for the same day. In our evaluation, although there is some consistency on several days (e.g., heart rate anomaly on 11/20), many outputs are scattered across the 24-hour period. Additionally, there are two significant long heart rate anomaly durations in the initial LLM outputs on 11/23 and 11/19 that are inaccurate. As a result, the LLM anomaly detection results not only misalign with the experts' mental model but are also highly inconsistent and unstable. This highlights the need for a more expert-instructed anomaly detection process to ensure greater stability.}

In the next section, we iterate our prototype based on the insights we got from the initial user study, and present our final prototype Vital Insight version 2 (\textbf{VI-2}).

\section{FINAL PROTOTYPE: SEAMLESS INTEGRATION OF AI ASSISTANCE WITH VISUALIZATION}
\begin{figure}[htbp]
  \centering
  \includegraphics[width=\linewidth]{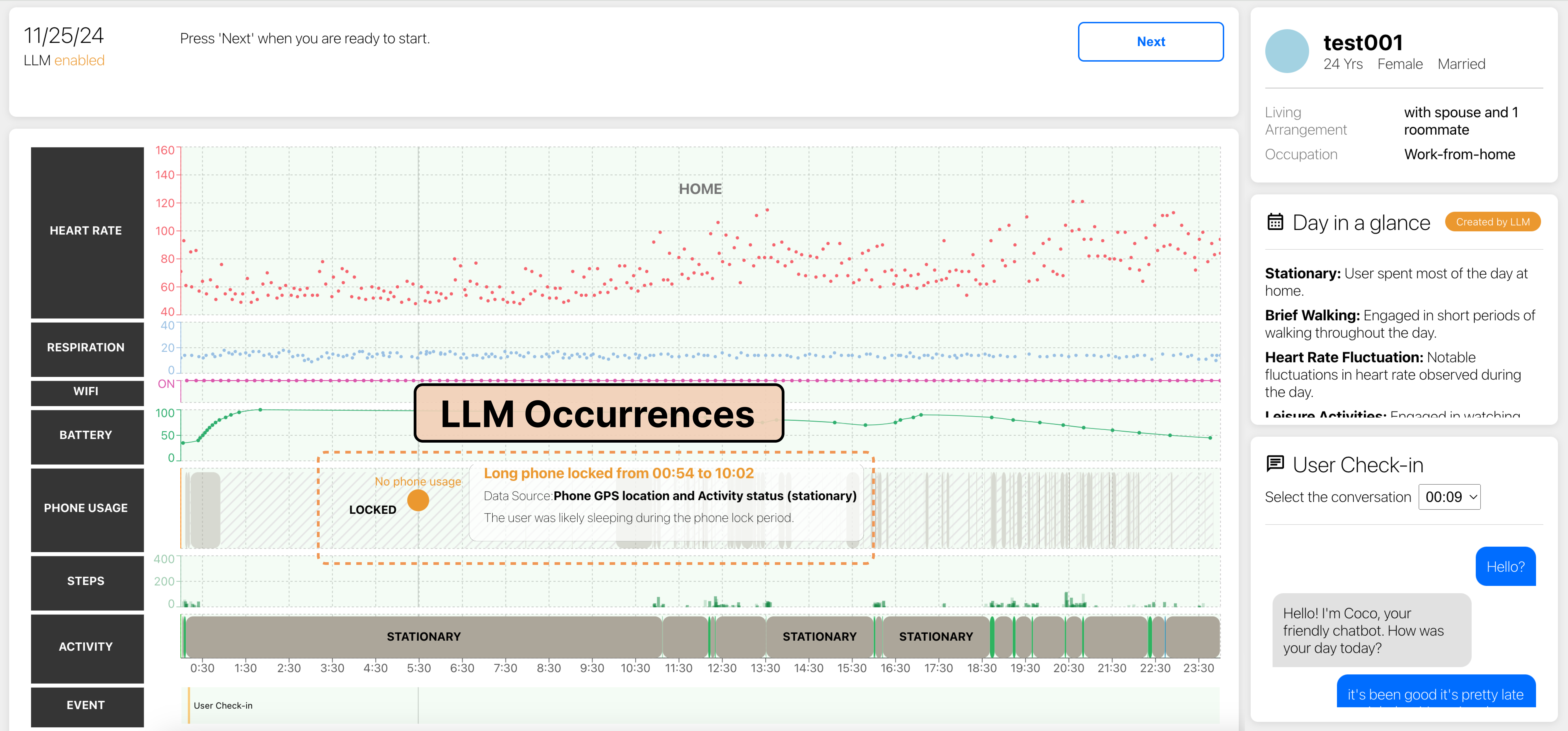}
  \caption{Design of the new iteration of Vital Insight (VI-2). Top left: user testing input; bottom left: main visualization and LLM detected occurrences; top right: user profile; middle right: Day in a glance generated by LLM; bottom right: user check-in conversation.}
  \label{fig:vi2}
  \Description{}
\end{figure}

Based on feedback from Study 2, a major area for improvement was the seamless integration of AI augmentation with the visualization.
For instance, experts expect not just a text-based summary, but the ability to extract important information and display it directly within the visualization to fill in the gaps.
Regarding trustworthiness, we found that experts struggled to trust the anomaly detection results, especially when they were purely generated by the LLM.
As a result, we modified the design to incorporate a more human-in-the-loop process to detect and explain important events, which we defined as ``occurrences'' in our final prototype.
Below, we detail the changes we made to design a final prototype, \textbf{Vital Insight 2 (VI-2)} (Figure. ~\ref{fig:vi2}).

We redesigned the user interface of \textbf{User Profile}, \textbf{User Check-in}, and \textbf{Day-in-a-Glance} (formerly the Daily Summary) components to be more intuitive and concise. 
For instance, we added timestamps for each conversation in the User Check-in and formatted the Day-in-a-Glance to create a clear design that highlights important activities in bold.
Additionally, we removed the Hourly Summary component, as experts rated it as the least helpful in the previous user study and led to additional cognitive load. Similarly, we also removed the Question component, which while interesting, did not directly assist the sensemaking process.
The \textbf{Main visualization} module received high ratings in Study 2, so we kept it largely unchanged in the next iteration barring some minor UI improvements like increasing the gap between each heart rate data point.

Furthermore, given the ambiguous definition and low trust in LLM-detected anomalies in Study 2, we now frame ``anomalies'' as significant \textbf{Occurrences} of interest.
Instead of letting the LLM directly detect anomalies without any guidance, we have now introduced a more \textbf{human-in-the-loop} process \jiachen{to mitigate potential inconsistencies and inaccurate results in LLM anomaly detection} (Figure. ~\ref{fig:llm}). 
In this approach, we used the experts' sensemaking process from Study 2 to first identify the important occurrences\jiachen{, which are typically preprocessed features. These features, along with the semantic descriptions of all modalities of sensing data around that time (as described in Section 4.3), are then sent to the LLMs for corresponding explanations.}
We identified five major occurrence features from Study 2: changes, long durations, discrepancy, gap, and routine activities.
For example, we observed that the changes in the location label was a key instance that caught the experts' attention, prompting them to begin their exploration and examination.
Thus, in VI-2 we detect all occurrences where the location label changed and sent the related data around that time period to the LLM to obtain explanations.

In addition to the backend improvements on how to identify and explain occurrences, VI-2 introduces a new way of displaying these occurrences to experts. 
Instead of being a separate module (like the anomalies module in VI-1), Occurrences and their LLM-generated explanations are now directly plotted on top of the related raw data streams, represented by orange dots with the name of that occurrence.
When a user hovers over an orange dot, they will see the title of the occurrence, a detailed inference generated by the LLM, and a description of the data sources used by the LLM to generate that inference.
This method represents an approach where, instead of providing a stand-alone result like a paragraph in Study 2, we aim to create AI assistance that is seamlessly integrated into the experts' sensemaking process.
The new LLM inferences support the non-chronological sensemaking process of experts as well as provide a source of input for AI assistance to build trust.
We envision that this newly updated prototype can better support the collaborative process between experts, data, and AI assistance.
We have made the system open-source and publicly accessible on GitHub at \url{https://github.com/UbiWell/vital-insight-public-repo}.

\section{STUDY 3: COMPARATIVE STUDY}
Since the main feedback from Study 2 centers around its AI component, in Study 3, we aimed to conduct a comparison study to assess whether the integrated AI assistance was truly helpful for experts' sensemaking tasks. 
We compared VI-2 with AI-assistance to a version of VI-2 without any AI-assistance.

\subsection{Method}
\subsubsection{Data Collection}
We used the same sensing system to collect data. 
Additionally, to determine whether the sensemaking results from researchers were accurate, participants self-reported fine-grained activities through ecological momentary assessments (EMA) multiple times a day, in addition to the daily check-ins with Alexa.
The EMA results typically consisted of a few words describing the current activity, such as ``going out for lunch.'' 
We used EMAs solely as the ground truth for momentary user behavior to determine the accuracy of if the sensemaking results from researchers. We did not provide the EMAs to the LLMs or the experts.

\subsubsection{Study Design}
We conducted an A/B test to investigate whether LLM assistance in VI-2 is helpful for researchers in performing sensemaking tasks. Condition A used VI-2 with full LLM assistance, while Condition B used VI-2 without LLM assistance. In Condition B, the experts did not have access to the Day-in-a-Glance and the Occurrences explanation.
During the experiment, each participant interacted with data from four different days, with two days for each condition, and answered a series of questions. 
The order of the tasks (A/B conditions and order of days) was randomized to ensure counterbalance and avoid a carry-over effect.
For each day, researchers first answered three multiple-choice questions related to the person’s day using all the information provided in the system. 
For example, a sample question could be ``What is the person doing at 6:30 pm?'', with several options like ``going out for lunch'', ``hanging out with friends'', etc.
After answering each question, experts rated their experience by completing the Single Usability Metrics survey (SUM) across the dimensions of ease, time, and satisfaction~\cite{sauro2005method}, along with two Likert-scale questions to rate the AI component (only in Condition A, where LLM is enabled).
Next, participants wrote a short summary of the person’s day within 3 minutes and completed the same surveys they filled out after answering the previous selection questions.
After each task, participants discussed their thought processes as qualitative feedback for the research team.
The questions module was integrated directly into the interface during the user testing ~\ref{fig:vi2}.
After the experiment, the participants filled out two System Usability Scale (SUS) to rate the overall usability of LLM-enabled and disabled prototypes~\cite{brooke1996sus}, and answered several questions related to the helpfulness of AI assistance.
At the end of the study, we asked participants several questions aimed at identifying potential actions based on the four days of data. 
For example, we asked what insights could be generated about the person, and inquired about the types of health interventions they might design to improve the person’s physical activity levels.
Since the correctness of the answers could not be objectively measured during this user testing, we kept this as an open-ended discussion with the experts to gather qualitative insights.

For Study 3, we followed the same recruiting and compensation process as Study 2. 
We recruited a total of 20 participants, with some overlap with Study 1 and 2 (Table. \ref{tab:demo}).
We recorded all user testing sessions and followed an open-coding approach to analyze the transcripts and notes as our qualitative analysis.
For the quantitative analysis, we calculated various metrics across the two conditions, including task completion time, correctness in answering the sensemaking questions by comparing with EMA ground truth, user-perceived metrics (satisfaction and ease), and the SUS score for the entire system.
\begin{figure}[htbp]
  \centering
  \includegraphics[width=\linewidth]{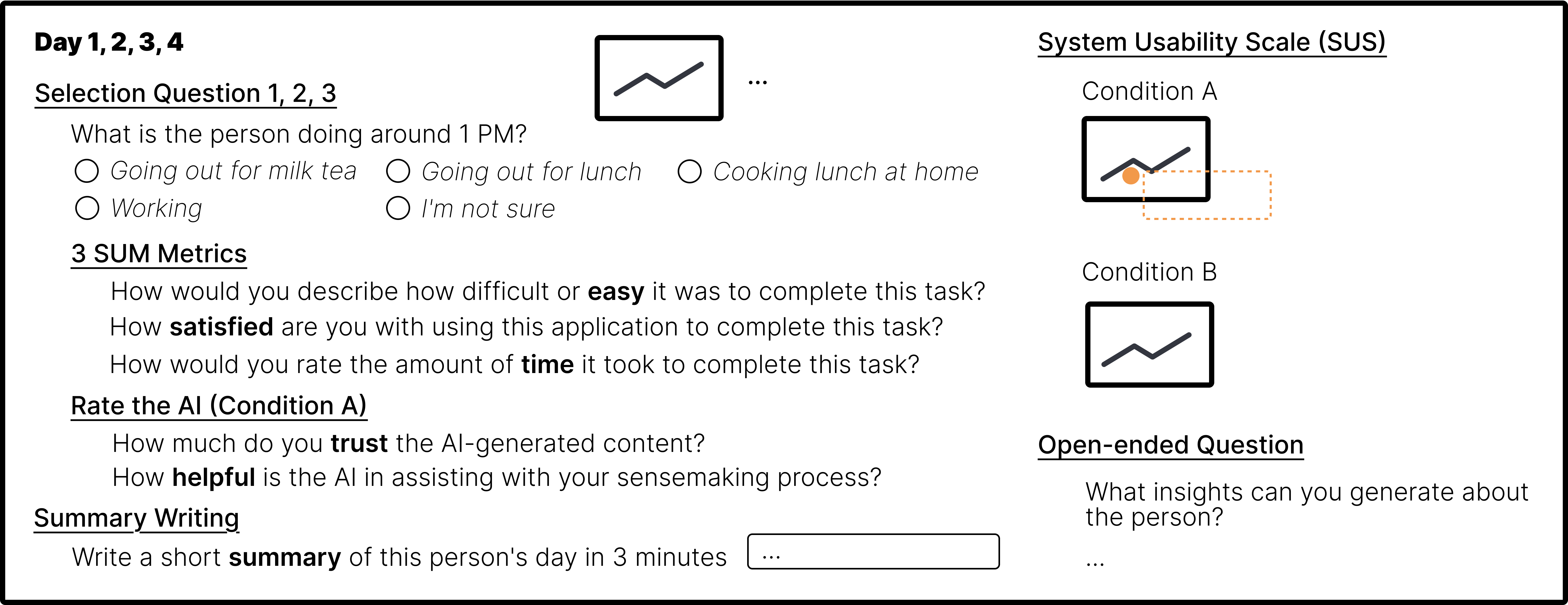}
  \caption{Study design of the A/B user testing (Study 3).}
  \label{fig:s3_design}
  \Description{}
\end{figure}
\subsection{Preliminary Evaluation on LLM Summary}
\jiachen{Although not the primary focus of our research, we conducted a preliminary evaluation of the LLM summary to assess its consistency, accuracy, and processing overhead to gain an overview of the system performance. We ran the LLM backend 10 times over data from four days and compared the LLM-generated summary with summaries written by multiple experts.}

\jiachen{For consistency, we calculated the similarity between summaries generated for the same day of data using the Term Frequency-Inverse Document Frequency (TF-IDF) method, combined with Cosine Similarity to avoid over-sensitivity to word frequency. We calculated the similarity matrix between each pair of LLM summaries (45 pairs for 10 summaries per day) and summaries written by different experts.}
\jiachen{The mean similarity of same-day summaries generated by LLM (Mean = 0.52, SD = 0.15) is significantly higher than those written by experts (Mean = 0.23, SD = 0.11), which aligns with the first author's observation. However, we want to emphasize that the summary of daily activities is highly subjective, and traditional NLP metrics like similarity might not perfectly capture the nuances of this type of task.}

\jiachen{Regarding accuracy, the first author manually reviewed each fact in the LLM and expert summaries, cross-referencing with passive tracking and EMA data. For example, the statement ``They appear to have gone to bed around 12:30 am and woke up around 9:15 am'' contains two facts: ``gone to bed around 12:30 am'' and ``woke up around 9:15 am.''
Each fact was categorized as `correct' if it aligned with self-report or sensing data, `wrong' if it contradicted the data, and `unclear' if the fact could not be conclusively determined from the available data.}
\jiachen{For the LLM summary, there were 5 wrong facts (2.13\%), 215 correct facts (91.49\%), and 15 unclear facts (6.38\%) out of 235 facts. For the expert summary, there were 15 wrong facts (11.03\%), 114 correct facts (83.82\%), and 7 unclear facts (5.15\%) out of 136 facts. LLM sometimes makes mistakes such as overstating details (e.g., ``Maintained a WiFi connection throughout the day'' when the WiFi was disconnected for a brief period), mismatches and hallucinations in time (e.g., ``Casual Shopping: User engaged in casual shopping activities, enhancing the evening's enjoyment.'' when shopping actually occurred in the morning). It also tends to draw overly aggressive connections between data (e.g., ``showing varied heart rate and respiration patterns while connected to WiFi'' when no clear relationship exists), and makes `reasonable' but slightly overstated and unnecessary guesses (e.g., ``The user frequently interacted with their phone throughout the day, suggesting engagement with social media, messaging, or other applications,'' despite no tracking of social media usage). Additionally, we observed an instance where the LLM had issues with numbers when summarizing heart rate data due to the loss of detailed raw data under the multi-level processing structure. For experts, many mistakes stem from missing data (e.g., ignoring details from user check-ins or profiles) or misinterpreting complex sensemaking between multiple modalities (e.g., wrongly identifying sleep when no steps are detected, but ignoring active phone usage data).} 

\jiachen{While the LLM summary had a higher accuracy rate, we found that expert summaries were generally more detailed, often including specific times, while LLM summaries tended to be more generalized about daily routines. For example, experts might summarize with ``They came back home around 7pm,'' while LLM might write ``Came back home later in the evening.'' We calculated the ratio of the number of facts included in the summary to the total number of tokens. The mean ratio was 5\% (SD = 1\%) for LLM-generated summaries and 9\% (SD = 2\%) for expert-generated summaries.  We will discuss more on the qualitative feedback from experts in the next section.}

\jiachen{Finally, we calculated the overhead for LLM usage. On average, there were 262182.25 input tokens and 1241.83 output tokens processed per day. For the GPT-4o-mini model used in Vital Insight, the API cost was \$0.0397 per day, approximately \$14.50 per year for a longitudinal deployment. Although the price is reasonable, it could increase if the system needs to run multiple times to avoid inconsistencies.}
\subsection{Results}
\subsubsection{Descriptive Results}
First, we discuss the descriptive results from the study. For correctness in answering the questions, we summed the scores for each participant to calculate a \textbf{final score}. For answers that were completely correct, we gave a score of 1; for those that were partially correct (e.g., missing one option in a multi-option question), we gave a score of 0.5; and for answers with any incorrect option, we gave a score of 0. Each participant answered 3 questions for each of the 4 days thus the final score ranged from 0 to 12. 
The final scores of 20  experts ranged from 1.5 to 11, with outliers (Median = 8.5, SD = 2.34). 
We believe there is no ceiling effect for the questions, and the study and question design are appropriate to capture some differences between the two conditions.
For each question where LLM was enabled, we asked participants to rate the \textbf{trustworthiness} and \textbf{helpfulness} of the AI component (scored 1=Distrust/Unhelpful to 5=Trust/Helpful),. 
The mean trustworthiness and helpfulness scores were 3.94 (SD = 0.80) and 4 (SD = 0.91), respectively, indicating a generally good experience with the LLM (4 = Somewhat Trustworthy/Helpful)\jiachen{, which aligns with other studies on AI-assisted systems that also consider a similar score to be a good rating ~\cite{yang2024talk2care,10296017}}.
\begin{figure}[htbp]
  \centering
  \includegraphics[width=\linewidth]{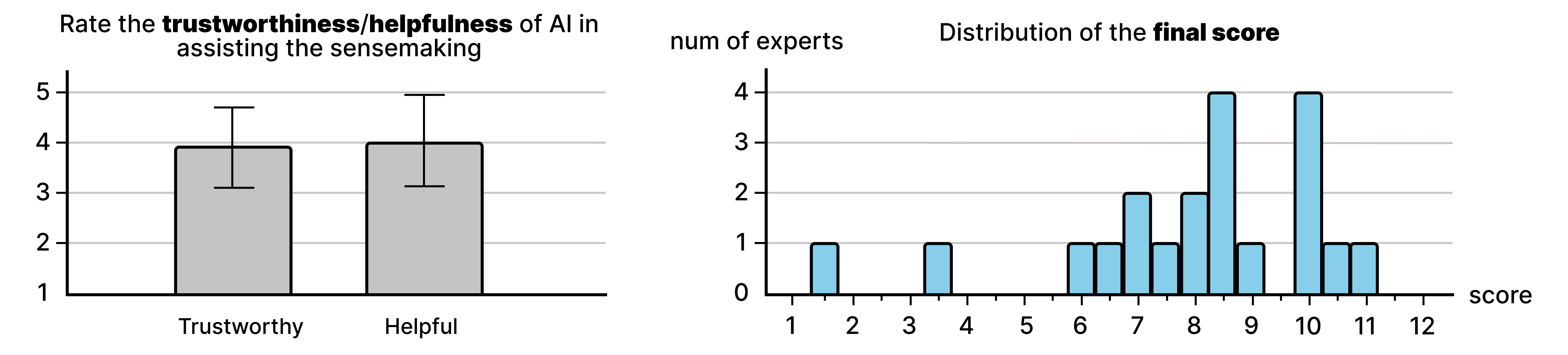}
  \caption{Descriptive results from the A/B user testing (Study 3).}
  \label{fig:des_s3}
  \Description{}

\end{figure}


\subsubsection{Results of A/B Test}
We additionally conducted statistical tests to evaluate the differences in completing sensemaking tasks (answering questions and writing summaries) with and without LLM. 
For ordinal variables, such as ease and satisfaction measured on a scale from 1 to 5, we used a Cumulative Logit Mixed-effects Model (CLMM) to examine the effect of LLM on participants' responses. For numerical variables, such as correctness (scored from 0 to 1) and completion time, we used Linear Mixed-Effects Models (LMER).
We included random intercepts for id (participant) and nested random effects for date and question, as each date had different questions, accounting for variability in responses within participants and across different dates and questions.
Since each participant was only exposed to one condition for the same date, we fitted two models: one that included an interaction term between condition and date to test whether the effect of condition varied across dates, and a simpler model that included only the main effect of condition.
We compared the models to assess whether the inclusion of the interaction term improved the model fit, using the Akaike Information Criterion (AIC) value and selected the model with lower AIC value.

We first measure the \textbf{correctness} of completing the task using a Linear Mixed-Effects Model (LMER).
We only consider the tasks of answering the selection questions (score = 0, 0.5, 1) for measuring correctness.
We did not find any significant difference between the two conditions in terms of correctness.
For the \textbf{ease} of completing the tasks (scale = 1-5), we used a Cumulative Logit Mixed-effects Model (CLMM).
We again found no significant difference between the two conditions regarding ease of task.
Similarly, we measured the \textbf{satisfaction} of completing each task using CLMM.
We found that users rated the AI-enabled version with significantly higher satisfaction scores than the non-AI version ($p<0.05$)
We then calculated the amount of \textbf{time} spent for each task in seconds from the screen recording, and analyzed the results using LMER.
We found that experts took approximately 14.72 more seconds to complete the task with AI content than without it ($p < 0.01$).

The distribution of \textbf{SUS scores} with LLM followed a normal distribution (mean = 73.42, SD = 17.92), while the distribution without LLM was negatively skewed (mean = 70.53, median = 75, SD = 19.59). 
The mean SUS scores were considered ``Good'' and ``Acceptable'' for the LLM-enabled condition, and ``OK'' and ``Marginal High'' for the LLM-disabled condition~\cite{bangor2009determining,bangor2008empirical}. 
Using a Wilcoxon Signed-Rank Test, we found that having AI-assistance led to significantly higher SUS scores than without AI-assistance($p < 0.05$), with a medium effect (effect size = 0.44).

To summarize the results, we did not find any significant quantitative differences between the AI and non-AI conditions in ease of task and correctness of the questions. 
However, there was a statistically significant difference in expert satisfaction and time spent on the tasks between the two conditions, with experts spending more time on the tasks and reporting higher satisfaction with AI. 
System usability was generally regarded as good and acceptable, with a significant difference in usability between the AI and non-AI conditions, favoring the AI-assisted condition.

\subsubsection{Interpretation of the Quantitative Results}
While interpreting the data, the difficulty of answering the questions depended on the specific question itself and the available data. 
This is reasonable, as the LLM has access to the same data that the expert users can retrieve from the prototype, and experts may simply be unable to answer certain questions with the given limited sensing data, regardless of the assistance provided.
Additionally, we did not limit the time for answering each question, allowing experts to use all available information to select the answer based on their knowledge of both conditions. 
This may have contributed to the lack of difference in correctness between the conditions.
The results also emphasize the importance of experts in making sense of the data, as well as the need to consider their trust in the LLM. 
We further discuss experts' interaction with the AI content through qualitative analysis in section 7.2.4.

While AI does not directly affect experts' sensemaking results, LLM does help experts in completing sensemaking tasks. 
This is supported by both the descriptive findings (helpfulness) and A/B testing results (satisfaction).
Experts also spent more time on the tasks when AI content was presented, which aligns with our qualitative findings, where they engaged in an iterative process to comprehend and validate the information.
In Section 7.2.5, we provide an in-depth discussion on how AI helps experts' sensemaking process.

\subsubsection{General Usability and Qualitative Feedback}
Many of the interactions and insights remain similar to Study 2, thus we focus on discussing only the new insights in this section.

Experts actively used both LLM occurrences and the day-in-a-glance feature. 
Furthermore, experts confirmed that LLM occurrences did their intended function of extracting important information and filling the gaps.
For LLM occurrences overlaid on the visualization, experts appreciated the temporal association, especially the details within specific timeframes that helped them answer certain questions (P13). 
The LLM was also significantly more helpful when there were multiple check-ins in a day or when the check-ins were particularly long.
\begin{quote}
    \emph{``I think LLM does a pretty good job of filling in the gaps by extracting necessary, relevant information at each time point. So even without going into too much detail, like a long conversation, I could kind of see a map between those two—like data and (user) feedback.'' (P19).}
\end{quote}
P15 confirmed that the AI component can help her save time when running longitudinal studies and identifying the quantity and quality of data.

Regarding trust in AI, experts seemed to trust the AI-generated content more than the first iteration because VI-2 explicitly describes the evidence used and the content aligns with that evidence.
While experts built more trust, they also started to ``ask for more'', compared to Study 2, where experts were more doubtful about the AI content. 
For instance, P17 mentioned liking one of the occurrence explanations because it provided an extensive description that combined multiple sensing data points to make a ``real'' inference, rather than simply extracting user check-in data.
This reflects something that wasn't fully revealed in Study 2: 
beyond merely verifying AI content, as users build trust in the AI, they treat it as an additional modality of information, and even use it to validate other information.
In one scenario, P13 observed a user check-in indicating that the person went out for milk tea and to the grocery store, but they still looked at the LLM-generated content to validate the information.
As experts gradually built trust in the LLM and started to view it as another reliable source of information, they spent more time comprehending and integrating its insights.

\subsubsection{User Interaction: Experts' Sensemaking Process}
In Study 3, we posed specific questions for experts to answer. This approach revealed new insights into the sensemaking process that are complementary to the more open-ended summary-writing scenario in Study 2. 
Some experts immediately began identifying related data based on the question. 
The first thing they tried to identify was time, which helped them narrow down the data set to focus on. 
Some questions also prompted experts to focus on a specific modality. 
For instance, when asked what the person was doing during a long phone usage period, many experts went directly to the phone usage data and only turned to other modalities as necessary.
In such cases, LLM occurrences became an \textbf{additional modality}, on par with other sensing modalities, assisting experts to make sense of what happened.
Interestingly, despite the specific questions, some experts still started by reviewing the entire day's data to first get an overview. 
This process is similar to what we observed in Study 2. 
In this scenario, both LLM occurrences and the day-in-a-glance feature were the first things some experts examined, especially when they were familiar with LLM and ``knew what to expect.''

The discussion on suggestions for high-level insights and actions led to some interesting findings. 
Compared to Study 2, where we showed only one day of data, displaying four days of data for the same person allowed experts to generate more longitudinal insights.
We noticed that although experts varied in their answers to specific questions, their high-level insights about the person remained remarkably consistent. 
For instance, all the experts concluded that the person led a sedentary lifestyle, and many believed the person was social.
We observed that a crucial sensemaking process was to \textbf{establish and enrich the user’s persona}. 
This persona included information not present in the contextual data provided and was more abstract than simple patterns, such as lifestyle, hobbies, social interactions, and regular routines. 
Experts then used these personas to suggest actions.
\begin{figure}[htbp]
  \centering
  \includegraphics[width=0.9\linewidth]{figs/model1.png}
  \caption{Experts employ detailed sensemaking procedures across various modalities, including sensing data, contextual information and AI contents to identify data patterns, construct personas, and determine potential actions.}
  \label{fig:sensemaking_data}
  \Description{Diagram illustrating sensemaking procedures with Sensing Data, Contextual Information, Pattern identification, AI-supported Inference, and Action. Sensing Data like heart rate and steps are compared and integrated into Patterns such as discrepancies and anomalies. Contextual Information, including user profile and location, helps extract or fill and explain patterns. AI-supported Inference covers inference, summary, and questions, interacting with Patterns to guide Actions like stakeholder outreach and debugging. Arrows indicate processes such as compare, integrate, extract or fill, and seek explanation, showing the flow between components. \jiachen{updated figure}}
\end{figure}
For example, they recommended using a standing desk due to the person’s work situation (P12/14), exercising socially with friends (P3/8/20), and incorporating more physical activity into daily routines like milk tea outings and dinner (P6/9/14/16), or considering a streaming exercise program since the person frequently watched streaming content (P8).
During this process, experts also integrated their own knowledge when deciding on actions, like P8 and P11 both suggested a gamified intervention.
\section{EXPERTS' SENSEMAKING MODEL}
In the previous section, we discussed the qualitative findings regarding the sensemaking processes employed by experts.
In this section, we will build on those findings, synthesizing them at a higher level to address \emph{RQ2: What is the sensemaking process experts employ when interpreting personal tracking data?}

Overall, experts' sensemaking process is iterative, continuously refining the previous decision when observing new data proof.
This process is usually a combination of chronological order and non-chronological order.
Throughout the whole process, switching between different modalities is a key sensemaking part and challenge.
Thus, we further scaffold the detailed sensemaking procedures between different data modalities employed by experts (Fig. \ref{fig:sensemaking_data}).

Experts frequently \emph{compare} various sensing modalities and \emph{integrate} them to synthesize patterns.
We identified five types of patterns that experts pay attention to: long duration, change, discrepancy, gap, and routine activities.
To explain these patterns, experts turn to contextual information like user profiles and check-ins, and use them to \emph{fill} in gaps that sensing data alone cannot provide, often involving a process of \emph{extracting} relevant details.
Based on the patterns, experts may take certain actions directly, such as addressing sensor issues. 
More commonly, however, experts work to \emph{synthesize} a user persona before taking action.
These personas are typically constructed from patterns observed in longitudinal data, along with known contextual information such as demographics. Common persona traits include lifestyle, hobbies, social interactions, and regular routines.
Once the persona is established, experts use it to guide specific actions, such as reaching out to relevant stakeholders or designing personalized interventions.
Furthermore, the sensemaking procedures described above can be supported by AI inference at varying levels of granularity. 
For instance, observation summaries of sensing data can help experts compare and integrate raw data across modalities to identify patterns, while more creative inferences can assist them in brainstorming potential explanations for those patterns.

\begin{figure}[htbp]
  \centering
  \includegraphics[width=0.7\linewidth]{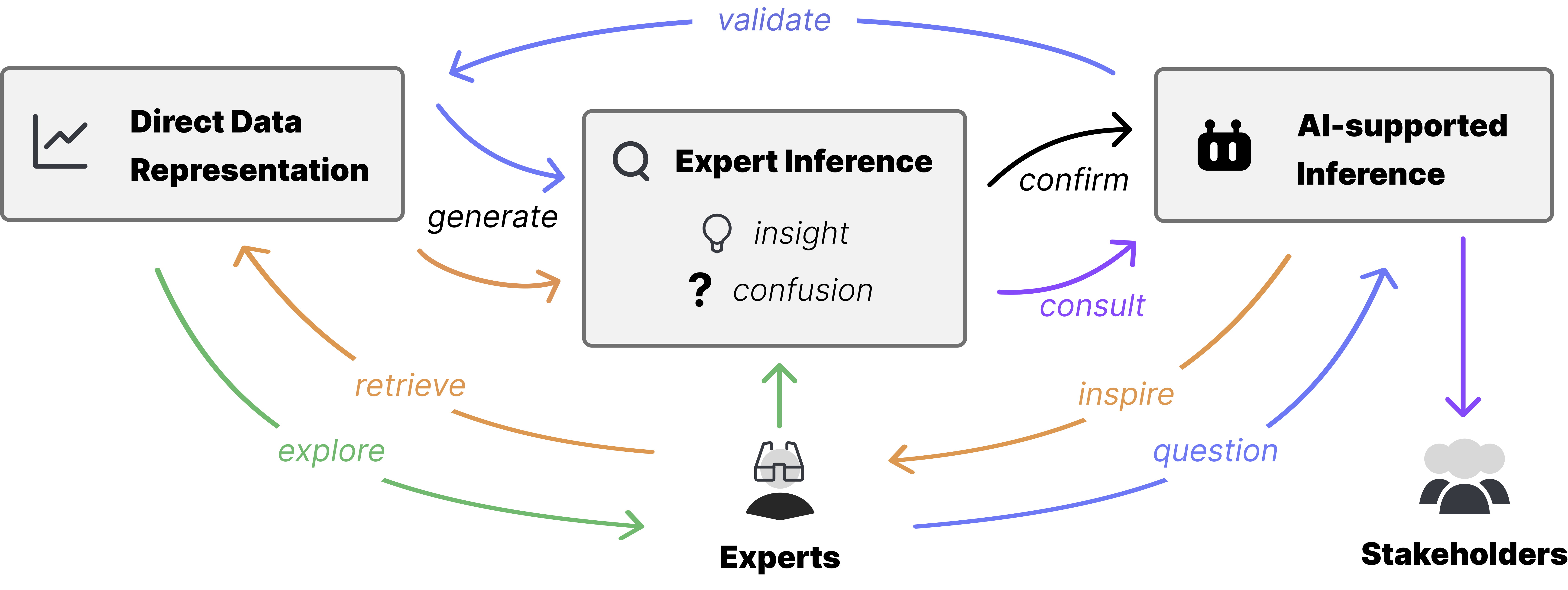}
  \caption{High-level experts' sensemaking procedure around direct data representation and AI-supported inference.}
  \label{fig:sensemaking_overall}
  \Description{Diagram of the high-level sensemaking procedure with Direct Data Representation, Expert Inference, and AI-supported Inference. Direct Data Representation on the left, expert inference in the middle with two subcomponents, insight and confusion, AI-supported inference on the right, with an expert icon under expert inference and stakeholder under AI-supported inference. Different components are connected through arrows. AI-supported inference `inspires' experts to `retrieve' data from direct data representation, then `generate' insights as expert inference. Experts `question' AI-supported inference, `validate' from direct data representation, and  `generate' insights as expert inference. Experts `explore' direct data representation, and `generate' insights as expert inference. There are also two arrows between expert inference and AI-supported inference, where experts `confirm' their insights from AI, and `consult' AI when there’s confusion. AI-supported inference also points to stakeholders icon.}
\end{figure}

To further elevate the sensemaking model, we consider the collaboration between experts, data, and AI.
Experts iteratively move between direct data representation and indirect inference to \textbf{generate} insights (Fig. \ref{fig:sensemaking_overall}). 
A common approach for experts is to \textcolor{DarkGreen}{explore} the data directly and develop their thoughts. 
Two other main paths involve indirect inference. 
One starts with \textcolor{blue}{questioning} certain aspects of an indirect inference and seeking evidence from data to \textcolor{blue}{validate} the results. 
The other involves being \textcolor{orange}{inspired} by AI-supported inference, such as overlooked details in indirect inference, and then intentionally \textcolor{orange}{retrieve} relevant information from the data. 
These three paths lead to insights, where experts feel confident in drawing conclusions. 
After reaching insights, some experts may revisit the indirect inference to \textbf{confirm} their assumptions.
In addition to generating insights, these paths may also lead to confusion. 
Some experts choose to \textcolor{violet}{consult} indirect inference for hints while others route to stakeholders like participants to reach confident answers.

\jiachen{Experts' sensemaking model also highlights the need for customization. While there are many commonalities in their mental models and thought processes, each expert's focus varies slightly. Differences include starting points, preferred modalities, the amount of data or time range they are comfortable reviewing, familiarity with the target population, and preferred processing modes (e.g., more visual or more text-based). Based on this model, the system should allow extensive customization across these aspects, including basic interface customization like layout and color following Schneiderman's mantra ``Overview first, zoom and filter, then details-on- demand''~\cite{shneiderman2003eyes}, as well as advanced customization based on the personal sensemaking model. It's also important to recognize the varying levels of trust in AI-generated content among different experts. Systems should tailor the tone of the AI to be more conservative, with additional proof and explanations for certain experts, while providing more insights from the LLM for those open to getting inspiration from the outputs.}

\section{DISCUSSION} 
\seclabel{discussion}
Building on our findings, we present design implications for systems designed to assist in the sensemaking process of personal tracking data, discuss contexts to help situate our findings, and offer suggestions for future directions.
\subsection{Design Implications}
In this section, we present a list of design implications along with specific suggestions to answer \emph{RQ3: What are the design implications for AI-augmented visualization systems aimed at assisting the sensemaking of personal tracking data?}
We limited the scope of the design implications to this specific format of assistance that aligns with the prototype to avoid overstating.
\subsubsection{Evidence-Based Sensemaking that Enables Validation to Build Trust}
The design of these tools must support evidence-based sensemaking, enabling a continuously validating and confirming process by skeptical experts. 
It should explicitly provide and identify the evidence used to generate indirect inferences, and allow experts to validate these findings using their own sensemaking methods.
This aligned with the challenges of self-calibration as outlined by Karahanoğlu et al.~\cite{Karahanoğlu_View}.
We further categorized 2 useful pieces of evidence in this scenario: the source of data being used to conclude the inference, and the explanation of the deduction procedure using these sources.
These two evidences highlights experts' need for AI assistance to have a human-like sensemaking process instead of machine-like strategies such as black-box in Machine learning.  
Moreover, the system should ensure a clear distinction between raw data and inferences to provide clear expectations to facilitate trust.

\subsubsection{Integration of Modalities to Assist Contextualization}
Tools must facilitate the seamless integration of various sensing data modalities and enable easy comparison of different modalities across the same scale. 
This capability enables experts to compare and analyze different data types simultaneously, offering a comprehensive view of an individual's behaviors. 
In our cases, time is the most important scale in making sense of multi-modal health sensing data.
As a result, visually aligning multiple raw sensing data modalities on a time scale instead of only showing a fragmented summary of a single modality is a helpful design decision to facilitate integration. 
We also noticed that other than raw sensing data that naturally aligns through time, contextual information such as self-checkin conversation and user profile are usually hard to align with the rest of the data.
This is also proven to be a main challenge in our user testing sessions.
As a result, an effective tool must provide and extract important contextual insights to help explain events. 
This includes integrating information about the individual, such as their user profile, proactive self-report data, and calculated features like location labels and baselines.
After extracting the information, aligning it back to the same scale of time is very helpful for experts' sensemaking process.
All these efforts help contextualize complex data to generate meaningful insights.
\subsubsection{Support an Iterative Sensemaking Process and Determining Actions}
The tool must support the iterative sensemaking process experts use, allowing them to continuously refine their analysis and interpretations as new findings emerge. 
This includes enabling easy switches between various levels of granularity by offering interactions like zooming in/out on the timescale and providing different layers of inference, such as hourly/daily insights and observational/creative interpretations. 
The sensemaking process also requires gathering input from various stakeholders, so the system should inspire experts with reasonable and actionable steps to engage stakeholders, e.g., offering sample self-report questions, suggestions for useful data to collect, and explanations or solutions for potential anomalies.
Furthermore, in longitudinal studies, it is crucial for the system to facilitate the construction of user personas from the data to help identify personalized solutions.

\subsubsection{Provide Both AI-Assisted Contextual Overview and Momentary Highlights}
Across our studies, we found that AI assistance should provide both a semantic overview and momentary highlights to facilitate the iterative sensemaking process.
The contextual overview, such as a daily summary, serves as the background context that is ``always present'', almost like an additional data modality that provides an overview of the person's day on a higher level. 
The momentary highlights, such as occurrences at specific times in VI-2, integrate well with other sensing data on the same time scale, and help experts quickly navigate important information.
Both functionalities are essential for experts’ sensemaking process, complementing and enhancing each other, and should be integrated in a successful application.

\subsection{Expert Sensemaking Model for Multi-Modal Health Sensing Datasets with AI Assistance: Commonalities, Variance, and Comparisons}
Past work has broadly defined sensemaking as the process of searching for a representation and encoding data within that representation to answer specific questions~\cite{russell1993cost}, or representing data in a schema that facilitates analysis~\cite{pirolli2005sensemaking}.
In our work, we verified that the thought processes employed by experts can be clearly defined as typical sensemaking.
To integrate our model with existing frameworks such as DIKW (Data, Information, Knowledge, Wisdom), we observed that experts' sensemaking begins with raw sensing entities (Data), progresses to identifying patterns (Information), and extends to generating potential explanations (Knowledge), which ultimately guide actionable steps (Wisdom)~\cite{Shaik_Tao_Li_Xie_Velásquez_2024}.
In our studies, we were able to capture each step and further scaffold the procedures. 
For instance, we recognized that two distinct types of data—sensing and contextual—require different sensemaking procedures. Additionally, we identified five patterns that are commonly important for experts: change, discrepancy, gap, long usage, and routine activities.
While the framework does not explicitly specify the role of AI as an assistant in the process, our findings verify that AI can provide valuable support at each stage of sensemaking.

To further examine various sensemaking processes and goals, Faisal et al. categorized six types of sensemaking representations~\cite{faisal2009classification}.
In the specific context of making sense of multi-modal health sensing data, experts primarily use argumentational representations, which are formed by integrating a series of claims to infer conclusions or establish inferential relationships.
Since health-sensing data related to human behavior is often time-based, many sensemaking processes fall under sequential representations. 
Chronologies, for example, are a common type of sequential representation. As noted in our findings, experts often examine data from beginning to end.
However, non-sequential representations frequently emerge iteratively, where human-in-the-loop AI augmentation can help experts quickly observe patterns.
Another category of sensemaking we observed is faceted representation, often used in abductive reasoning, where humans try to derive the most reasonable explanation from the data. The authors describe this process as finding or adding different (often new) entities that influence sensemaking~\cite{faisal2009classification}.
In our study, the entities—sensing modalities—are fixed, meaning researchers theoretically cannot acquire additional data. However, this process still applies from a different perspective: the abundance of modalities can make it challenging to become familiar with all of them at once.
Experts typically start with a subset of modalities and gradually explore additional information as needed. Beyond modalities, ``the place to start'' also involves what we defined as ``occurrences'' in Study 3 -- specific events that immediately draw attention.
Another element that could be considered a facet is AI assistance, which can be viewed as an augmentation of data and visualization, as described by Wu et al.~\cite{wu2021ai4vis}.
In our study, particularly in Study 3, AI assistance occasionally transcends mere data augmentation, becoming a unique facet of the data source within experts' sensemaking models. For example, after becoming familiar with the system, many experts prioritize the AI's daily summaries or explanations of occurrences to guide their sensemaking process. They may then refer back to the raw data, rather than first examining the data and consulting AI content only when necessary. These interactions were described as the ``inspire-retrieve'' route in Fig. \ref{fig:sensemaking_overall}.
Although AI does not generate new data, it is crucial to recognize its standalone role as a new facet of information within human-in-the-loop sensemaking processes.

To summarize, our findings expand the existing understanding of experts' sensemaking models. This includes detailing specific procedures for interpreting multi-modal health sensing data and incorporating the role of AI assistance into the framework.

\subsection{Biases in LLM-Generated Inferences}

During user testing with experts, we observed them gradually building trust in the LLM-generated inferences and summaries while exploring the various system components.
However, this process also increased the potential risk of bias introduced by inaccurate LLM results. 
Several prior works have discussed concerns with bias in LLM-generated outcomes: highlighting biases related to gender~\cite{wan2023kelly}, age~\cite{duan2024large}, geography~\cite{manvi2024large}, and politics~\cite{urman2023silence}. 
These biases could influence the interpretation of contextual information about individuals. 
For example, LLM may offer a stereotyped interpretation of the data based on the user's certain demographics.
LLMs also exhibit bias in their information retrieval processes, such as giving preference to documents or items from specific input positions like the beginning or end of the list~\cite{dai2024neural,li2023halueval,dai2024bias}.
This bias could potentially affect the retrieval of various raw sensing data.
Since our user testing primarily involved people with extensive expertise, they expressed a highly skeptical view of the LLM-generated results, often validating the inferences before accepting them. 
For example, P2 specifically mentioned concerns about being influenced by the LLM summary and stated they would avoid checking it first to prevent bias.
However, as their trust in the system grew, there is a risk that they might be increasingly influenced by the LLM’s outputs, even for experts~\cite{choudhury2024large,kerasidou2022before,ryan2020ai}. 
During our internal evaluation of the LLM-generated results, we also noted a potential concern with over-interpreting sensor data, where the LLM might notice trivial fluctuations in heart rate and force itself to make an inference. 
Without explicit instructions, LLMs also tend not to question the validity and reliability of the data and always offer a potential explanation regardless.
\jiachen{While experts may be able to identify incorrect outputs from the LLM, there could still be a subtle bias introduced. Even without considering incorrect outputs, the possibilities provided by the LLM can introduce a small bias in what experts perceive as 'what might have happened.'}
Future research must recognize the importance of examining both the LLM's biases in interpreting personal tracking data and the biases experts might develop when integrating those results.

\subsection{Designing LLM Agents that Mimic Experts' Sensemaking Process}
Although this study did not aim to enhance the accuracy of complex activity recognition and sensemaking, our insights into experts' sensemaking processes may help inform the design of more effective AI-assisted systems for interpreting personal tracking data.
Guiding LLMs with expert knowledge is a common approach for solving domain-specific problems, including techniques like adding instructions in prompts~\cite{ji2024hargpt,cosentino2024towards,choube2025gloss}, creating expert agents~\cite{yang2024drhouse}, and fine-tuning~\cite{kim2024health,cosentino2024towards}.
In the context of making sense of sensor data, HARGPT employs simple role-playing prompts without expert guidance to identify activities from raw IMU data~\cite{ji2024hargpt}.
Health-LLM evaluated zero/few-shot learning and fine-tuning with simple prompts on sensor data~\cite{kim2024health}.
Similarly, prior works have also explored how to encode sensor data to prompts: LLMSense and PhysioLLM use text-formatted sensor data and prompts to derive high-level inferences from sensor traces~\cite{ouyang2024llmsense,fang2024physiollm}; Cosetino~et~al. incorporated domain knowledge into prompts and fine-tuned models using expert responses~\cite{cosentino2024towards}.
We build on previous works to provide an iterative and evolving approach to continuously include domain expertise as Data Interpretation Guidance in the prompt.
While prior works have explored incorporating domain/expert knowledge, we observed that much of the expert knowledge is usually ``static.''.
From related work, it is clear that the complex sensemaking process experts use when interpreting personal tracking data is under-documented and rarely considered to guide LLMs. 
Our study reveals that these sensemaking processes extend beyond basic knowledge, like the normal range of a person's heart rate, and encompass the selection and integration of multiple data streams, handling data discrepancies and uncertainties, and strategies for information retrieval.
These domain-specific insights can guide LLMs in various ways. 
For instance, understanding how experts initiate sensemaking -- starting from changes rather than following a strict chronological order -- could be used to optimize data retrieval for LLMs. 
Additionally, the differing definitions of anomalies between the LLM and experts highlight an additional area for improvement, involving extending the system to incorporate more task-specific and sophisticated anomaly detection algorithms.
Our findings on how experts collaborate with LLMs and other informational modules to reach conclusions underscore how LLM detection models should be designed for real-world applications to support experts effectively.
We hope that, although not the main focus of this study, our insights will help AI and LLM researchers recognize opportunities to enhance current detection systems and better support complex sensemaking tasks.

\subsection{Replacing or Assisting: Experts and AI Collaboration in Contextual Reasoning of Sensing Data}
\jiachen{Prior work using LLMs for contextual reasoning and sensor data understanding has predominantly focused on detection tasks -- building AI systems that make inferences \textit{independently} and compare their performance against human experts. The underlying goal in these cases is often to replace or reduce reliance on experts by demonstrating comparable or superior accuracy. In contrast, our work emphasizes a complementary vision where AI systems are designed to \textit{assist} experts with \textit{contextual reasoning} and \textit{understanding}: a more interpretive and subjective process. We frame LLMs as collaborative partners that assist experts in synthesizing multimodal, complex, subjective, and context-dependent data, rather than simply predicting predefined outcomes. This distinction is critical, particularly given the inherent challenges in sensemaking such as ambiguity, trust, and cognitive bias.
We recognize the importance of both goals and, in this section, we further discuss the different approaches taken in previous studies and our work to help guide future researchers in applying appropriate methods.}

\jiachen{Previous works comparing LLM results with human input often focus on tasks with well-defined outcomes, such as activity recognition, stress scores, or diagnoses~\cite{yang2024drhouse,ji2024hargpt,kim2024health,cosentino2024towards,ouyang2024llmsense,fang2024physiollm}. These tasks allow researchers to define clear metrics and conduct quantitative evaluations to claim success of certain techniques proposed in those studies, like prompt engineering techniques, multi-agent pipeline, fine-tuned model, and more.
However, for more subjective tasks like summarizing a person's day, there is no single ``right'' or ``wrong'' answer, thus making objective, quantitative comparisons difficult. 
For example, we used standardized NLP methods to measure the accuracy and consistency of the summaries, and found that expert-written summaries were significantly more diverse than those generated by LLM. 
However, these metrics are not enough to claim that LLM behaves better than experts.
In real-world studies, both types of results are necessary for different applications. Comparative studies provide valuable insights for building fundamental models that support applications requiring automated outputs. For instance, activity recognition results generated by LLMs can be directly used as context for delivering personalized health interventions.
Subjective results, on the other hand, are crucial for human-centered applications that involve multiple stakeholders, such as monitoring dashboards for researchers, daily summaries for self-reflection, or even clinical decision making.
Adler et al., in their recent paper, proposed the concept of actionable sensing as an alternative to detection research, which brings passive sensing data on behavior and physiology into clinical encounters to improve patient care~\cite{adler2024beyond}. We embrace a similar concept, where LLMs' extensive narrative capabilities create new opportunities for generating high-level content that is valuable for various stakeholders, extending beyond detection. 
}

\jiachen{The techniques and applications of LLMs for the two goals are also different. In comparison studies, the focus is often on generating an accurate final result, whereas in AI aiming to assist experts, much of the value lies in providing explanations and building trust. As a result, other than approving the accuracy, ensuring alignment with experts’ mental models becomes a critical design implication for the LLM pipeline of these systems. In this paper, one of our key contributions is the experts’ sensemaking model of sensing data, where we suggest that future AI-assisted systems provide outputs that follow and support this process. We define several tasks where experts may need assistance in making sense of the data, and instruct LLMs to provide the desired information along with proper explanations. In works that focus on comparing LLM and expert outputs, experts’ sensemaking models or similar insights are often used as external knowledge resources to guide the LLM in making better decisions through prompt engineering, fine-tuning, and reinforcement learning.}

\jiachen{To summarize, both goals of ``replacing with'' and ``assisting'' experts are valuable directions for future research; however, the latter, \textit{specifically related to enhancing understanding and contextual reasoning}, has generally received less attention in the past.
Recently, more and more researchers have recognized the significance of human-AI collaboration in understanding data, particularly within the healthcare context.
For instance, recent work by Zhang et al. proposed a new clinician-AI collaboration paradigm instead of an existing human-AI competition
paradigm to account for the high uncertainty and risks while diagnosing sepsis~\cite{zhang2024rethinking}.
We hope our work aligns with this direction and encourages more researchers to recognize the value and potential of AI in supporting experts in making sense of sensing data.
}

\subsection{Socio-Technical Gaps in Real-world Practices: Stakeholders, Privacy and More}
In this study, the primary audience for our system is experts in multi-modal sensing data. 
However, we recognize that the ultimate goal of sensemaking often involves various stakeholders, including caregivers, family members, and clinicians, especially in health-related scenarios like elderly care or recovering from substance use disorder. 
This goes beyond merely gathering information from different stakeholders to interpret the data, but also includes enabling them to use the data in meaningful ways and take necessary actions. 
Upon closer examination of real-life practices, we realized that these efforts often transcend data sensemaking and involve addressing broader socio-technical gaps in ubiquitous computing~\cite{ackerman2000intellectual}.
For instance, sharing personal tracking data with caregivers and family members is not simply a matter of asking for ``the best recognition result of the current activity.'' 
Instead, it’s a much more complex process involving privacy, agency, trust, utility, and the dynamics of care~\cite{li2023privacy,townsend2021,schomakers2022privacy}.
Although we did not design the current version of our prototype for use by other stakeholders, we still discuss the opportunities and challenges of presenting such information if our system were to be extended to other stakeholders.

When presenting data to stakeholders with less expertise in sensing, the visualization needs to be aggregated into simpler, high-level information~\cite{li2023privacy,mynatt2001digital}. 
The sensemaking process for stakeholders like caregivers might also be action-driven more than insight-driven. 
Instead of moving from inference to evidence, their process often follows a pattern of identifying necessary actions first~\cite{wherton2008technological}. 
Therefore, notifications such as alarms and reminders may be more effective, or simply a ``peace of mind'', reassurance that everything is normal and no extra attention is required~\cite{mynatt2001digital,li2023privacy}. 
Regarding privacy, past research has shown that people are generally more comfortable sharing sensor data than conversations or stress indicators~\cite{raij2011privacy}. 
However, this acceptance may be based on the perception that only limited insights can be derived from raw data. 
When experts and systems can extract more information than users anticipated, it becomes crucial to provide proper privacy education and a comprehensive consent process, underscoring the need for more human-centered approaches to determining what information should be disclosed.

\subsection{Limitations}
The study has several limitations worth noting. 
While we achieved data saturation, our small sample size limits the generalizability of our findings, particularly as it relates to other health and behavioral outcomes. 
Although recent research suggests small sample sizes are acceptable in interpretive studies~\cite{crabtree2024h}, future work should involve larger populations and diverse scenarios to validate the findings. 
While researchers reviewed all the LLM-generated content shown in the two user studies to ensure basic correctness without making any changes, we \jiachen{conducted a preliminary quantitative evaluation of the LLM results}. 
A more thorough evaluation
of LLM results and capabilities can be conducted in future research, which is beyond the scope of this study.
Another limitation arises from the fact that experts spent more time on tasks when LLM content was included compared to when it was not. 
In our qualitative analysis, we found that this was partly due to experts treating LLM content seriously in their sensemaking process, which led them to spend more time on it.
Additionally, although we provided a tutorial video at the beginning, there was still a learning curve in using the information effectively.
A more longitudinal study is needed to assess the long-term usage of the prototype in real-world settings and determine whether it can ultimately save experts' time in sensemaking tasks.
\jiachen{Although we conducted A/B testing using a baseline system without LLM, we did not compare our system with other dashboards due to the specificity of the task. As a result, we presented the survey results as descriptive measurements and highlighted the positive qualitative feedback from experts, which aligned with the survey findings. }



\section{Conclusion} 
\seclabel{conclusion}

In this study, we conducted human-centered sessions with 21 experts to understand how we can support their sensemaking process in interpreting personal tracking data to support health and behavioral outcomes.
We conducted interviews and designed and developed two iterations of Vital Insight, a prototype that provides both direct data representation and indirect inference through visualization and Large Language Models (LLMs). 
We evaluated our prototype in user testing sessions with experts and presented a list of design implications.
By observing experts interacting with our prototype, we synthesized an experts' sensemaking model. 



\ifanonymized
 \relax
\else
 \section*{Acknowledgments}
 This research is partially supported by
the NIH National Institutes of Health, under award number
  P30DA029926, 
 and the National Science Foundation, under award number CAREER 2442593, 
 and 2112633, 
 The views and conclusions contained herein are those of the authors and should not be interpreted as necessarily representing the official policies, either expressed or implied, of the sponsors.
 Any mention of specific companies or products does not imply any endorsement by the authors, by their employers, or by the sponsors.
\fi 



\bibliographystyle{plain}	

\bibliography{main}
\end{document}